\documentclass[preprint,12pt]{elsarticle}

\usepackage{amssymb}

\journal{XXX}

\begin{document}

\begin{frontmatter}



\title{An explicit representation and enumeration for self-dual cyclic codes
over $\mathbb{F}_{2^m}+u\mathbb{F}_{2^m}$ of length $2^s$}


\author{Yuan Cao$^{a, \ b}$, Yonglin Cao$^{a, \ \ast}$, Hai Q. Dinh$^{c, \ d}$, Somphong Jitman$^{e}$}

\address{$^{a}$School of Mathematics and Statistics,
Shandong University of Technology, Zibo, Shandong 255091, China\\

\vskip 1mm $^{b}$ School of Mathematics and Statistics, Changsha University of Science and Technology,
Changsha, Hunan 410114, China

\vskip 1mm $^{c}$Division of Computational Mathematics and Engineering, Institute for Computational
       Science, Ton Duc Thang University, Ho Chi Minh City, Vietnam

\vskip 1mm $^{d}$Faculty of Mathematics and Statistics, Ton Duc Thang University, Ho Chi Minh City,
      Vietnam

\vskip 1mm $^{e}$Department of Mathematics, Faculty of Science, Silpakorn University, Nakhon
Pathom 73000, Thailand}
\cortext[cor1]{corresponding author.  \\
E-mail addresses: yuancao@sdut.edu.cn (Yuan Cao), \ ylcao@sdut.edu.cn (Yonglin Cao),
\ hdinh@kent.edu (H. Q. Dinh), \ sjitman@gmail.com (S. Jitman).}

\begin{abstract}
Let $\mathbb{F}_{2^m}$ be a finite field of cardinality $2^m$ and $s$ a positive integer.
Using properties for Kronecker product of matrices and calculation for linear equations over $\mathbb{F}_{2^m}$,
an efficient method for the construction of all distinct self-dual cyclic codes  with length $2^s$ over the finite chain ring $\mathbb{F}_{2^m}+u\mathbb{F}_{2^m}$
$(u^2=0)$ is provided. On that basis,
an explicit representation for every self-dual cyclic code  of length $2^s$ over $\mathbb{F}_{2^m}+u\mathbb{F}_{2^m}$
and an exact formula to count the number of all these self-dual cyclic codes are given.
\end{abstract}

\begin{keyword}
Cyclic code; Self-dual code; Linear code; Finite chain ring;
Kronecker product of matrices; Linear equations

\vskip 3mm
\noindent
{\small {\bf Mathematics Subject Classification (2000)} \  94B15, 94B05, 11T71}
\end{keyword}

\end{frontmatter}


\section{Introduction}

\noindent
  The class of self-dual codes is an interesting topic in coding theory duo to
their connections to other fields of mathematics such as Lattices, Cryptography, Invariant Theory, Block designs, etc.
A common theme for the construction of self-dual codes is the use of a computer search. In order to make this search feasible, special construction methods have been used to reduce the search field.
In many instances, self-dual codes have been found by first finding a code over a ring and then mapping
this code onto a code over a subring through a map that preserves duality. In the literatures, the mappings typically map
to codes over $\mathbb{F}_2$, $\mathbb{F}_4$ and $\mathbb{Z}_4$ since codes over these rings have had the most use.

\par
   Let $\mathbb{F}_{p^m}$ be a finite field of $p^m$ elements, where $p$ is a prime number, and denote
$R=\frac{\mathbb{F}_{p^m}[u]}{\langle u^2\rangle}=\mathbb{F}_{p^m}
+u\mathbb{F}_{p^m} \ (u^2=0).$
Then $R$ is a finite chain ring and every invertible element in $R$ is of the form: $a+bu$, $a,b\in \mathbb{F}_{p^m}$ and $a\neq 0$.
Let $N$ be a fixed positive integer and
$$R^N=\{(a_0,a_1,\ldots,a_{N-1})\mid a_0,a_1,\ldots,a_{N-1}\in R\}$$
be an $R$-free module with the usual componentwise addition and scalar multiplication by elements of $R$.
Then a nonempty subset $\mathcal{C}$ of $R^N$ is an \textit{$R$-submodule} of $R^N$ if
$\xi+\eta, a\xi\in \mathcal{C}$, for all $\xi,\eta\in \mathcal{C}$ and $a\in R$.
Let $\gamma$ be an invertible element in $R$. The \textit{$\gamma$-cyclic shift operator} $\sigma_\gamma$ on
$R^N$ is defined by
$$\sigma_\gamma(a_0,a_1,\ldots,a_{N-2},a_{N-1})=(\gamma a_{N-1},a_0,a_1,\ldots,a_{N-2}),$$
for all $(a_0,a_1,\ldots,a_{N-2},a_{N-1})\in R^N$. Then $\sigma_\gamma$ is
an invertible $R$-linear transformation on $R^N$. An $R$-submodule $\mathcal{C}$ is said
to be \textit{$\sigma_\gamma$-invariant} if
$$\sigma_\gamma(a_0,a_1,\ldots,a_{N-2},a_{N-1})\in \mathcal{C}, \ \forall (a_0,\ldots,a_{N-2},a_{N-1})\in \mathcal{C}.$$

\par
  In coding theory, an
$R$-submodule of $R^N$ is called a \textit{linear code} over
$R$ of length $N$ and every $\sigma_\gamma$-invariant $R$-submodule of of $R^N$ is
called a \textit{$\gamma$-constacyclic code}.
In particular, a $\gamma$-constacyclic code $\mathcal{C}$ is called a \textit{negacyclic code} when $\gamma=-1$, and
$\mathcal{C}$ is called a \textit{cyclic code} when $\gamma=1$.

\par
   Let $\frac{R[x]}{\langle x^N-\gamma\rangle}=\{\sum_{i=0}^{N-1}a_ix^i\mid a_0,a_1,\ldots,a_{N-1}\in R\}$
in which the arithmetic is done modulo $x^N-\gamma$.
In this paper, $\gamma$-constacyclic codes over
$R$ of length $N$ are identified with ideals of the ring $\frac{R[x]}{\langle x^N-\gamma\rangle}$, under the
identification map $\theta: R^N \rightarrow \frac{R[x]}{\langle x^N-\gamma\rangle}$ defined by
$\theta: (a_0,a_1,\ldots,a_{N-1})\mapsto
a_0+a_1x+\ldots+a_{N-1}x^{N-1}$ for all $a_i\in R$ and $i=0,1,\ldots,N-1$.

\par
  The \textit{Euclidean inner
product} on $R^N$ is defined by $[\alpha,\beta]=\sum_{i=0}^{N-1}a_ib_i\in R$ for
all $\alpha=(a_0,a_1,\ldots,a_{N-1}), \beta=(b_0,b_1,\ldots,b_{N-1})\in R^N$. Then
the (Euclidean) \textit{dual code} of a linear code $\mathcal{C}$ over $R$ of length $N$ is defined by
$$\mathcal{C}^{\bot}=\{\beta\in R^N\mid [\alpha,\beta]=0, \ \forall \alpha\in \mathcal{C}\},$$
which is also a linear code over $R$ of length $N$. In particular, $\mathcal{C}$ is said to be
 (Euclidean) \textit{self-dual} if $\mathcal{C}^{\bot}=\mathcal{C}$.

\par
  Let $\alpha=a+bu\in R$ where $a,b\in \mathbb{F}_{2^m}$. As in [3], we define
$\phi(\alpha)=(b,a+b)$ and define the \textit{Lee weight} of $\alpha$ by
${\rm w}_L(\alpha)={\rm w}_H(b,a+b)$, where ${\rm w}_H(b,a+b)$ is the Hamming weight of the vector
$(b,a+b)\in \mathbb{F}_{2^m}^2$. Then $\phi$ is an isomorphism of $\mathbb{F}_{2^m}$-linear
space from $R$ onto $\mathbb{F}_{2^m}^2$, and can be extended to an isomorphism of $\mathbb{F}_{2^m}$-linear
spaces from $\frac{R[x]}{\langle x^{N}-1\rangle}$ onto $\mathbb{F}_{2^m}^{2N}$ by the rule:
$$
\phi(\xi)=(b_0,b_1,\ldots,b_{N-1},a_0+b_0,a_1+b_1,\ldots,a_{N-1}+b_{N-1}),
$$
for all $\xi=\sum_{i=0}^{N-1}\alpha_ix^i\in \frac{R[x]}{\langle x^{N}-1\rangle}$, where $\alpha_i=a_i+b_iu$ with $a_i,b_i\in \mathbb{F}_{2^m}$ and $i=0,1,\ldots,N-1$.

\par
  The following conclusion is derived from Corollary 14 of [3]:
   \textit{Let $\mathcal{C}$ be an ideal of $\frac{R[x]}{\langle x^{N}-1\rangle}$ and set $\phi(\mathcal{C})=\{\phi(\xi)\mid \xi\in \mathcal{C}\}\subseteq \mathbb{F}_{2^m}^{2N}$. Then}

\begin{description}
\item{(i)}
   \textit{$\phi(\mathcal{C})$ is a $2$-quasi-cyclic code over $\mathbb{F}_{2^m}$ of length $2N$}.

\item{(ii)}
 \textit{The Hamming weight distribution of $\phi(\mathcal{C})$ is exactly the same as the Lee weight distribution of $\mathcal{C}$}.

\item{(iii)}
 \textit{$\phi(\mathcal{C})$ is a self-dual code over $\mathbb{F}_{2^m}$ of length $2N$ if
$\mathcal{C}$ is a self-dual code over $R$ of length $N$}.
\end{description}

\par
  Therefore, it is an effective way to obtain self-dual and $2$-quasi-cyclic codes over $\mathbb{F}_{2^m}$ of length $2N$
from self-dual code over $R$ of length $N$.


\par
  There were a lot of literatures on linear codes, cyclic codes and
constacyclic codes of length $N$ over rings $\mathbb{F}_{p^m}+u\mathbb{F}_{p^m}$ ($u^2=0$) for various prime $p$ and positive integers $m$ and some positive integer $N$.
For example, \cite{s1}--\cite{s4}, \cite{s6}--\cite{s17} and \cite{s19}.
The classification of self-dual codes plays an important role in studying their structures and encoders.
However, it is a very difficult task in general, and only some codes of special lengths over certain finite
fields or finite chain rings are classified.

\par
  For example,
all constacyclic codes of length $2^s$ over the Galois extension
rings of $\mathbb{F}_2 + u\mathbb{F}_2$ was classified and their detailed structures was also established in \cite{s8}. Dinh \cite{s9}
classified all constacyclic codes of length $p^s$ over $\mathbb{F}_{p^m}+u\mathbb{F}_{p^m}$.
    Dinh et al. \cite{s10} studied
negacyclic codes of length $2p^s$ over the ring $\mathbb{F}_{p^m}+u\mathbb{F}_{p^m}$.
Chen et al. \cite{s7} investigated
constacyclic codes of length $2p^s$ over $\mathbb{F}_{p^m}+u\mathbb{F}_{p^m}$.
Dinh et al. \cite{s11} studied constacyclic codes of length $4p^s$ over $\mathbb{F}_{p^m}+u\mathbb{F}_{p^m}$ when $p^m\equiv 1$ (mod $4$).
These papers mainly used the methods in \cite{s8} and \cite{s9}, and the main results and their proofs  depend heavily on the code lengths $p^s$, $2p^s$ and $4p^s$. It is particularly important to note that \textsl{the representation and enumeration for self-dual cyclic codes were not studied in these papers}.

\par
   Dinh et al. \cite{s12} determined the algebraic structures of all cyclic and negacyclic codes
of length $4p^s$ over $\mathbb{F}_{p^m}+u\mathbb{F}_{p^m}$, established the duals of all such codes and given some subclass of self-dual negacyclic codes of length $4p^s$ over $\mathbb{F}_{p^m}+u\mathbb{F}_{p^m}$ by Theorems 4.2, 4.4
and 4.9 of \cite{s12}. But \textsl{the representation and enumeration for all self-dual negacyclic codes and all self-dual cyclic codes
were not obtained}.

\par
  In Section 4 of \cite{s5}, Choosuwan et al. given
an alternative characterization for cyclic codes over $\mathbb{F}_{p^k}+u\mathbb{F}_{p^k}$
of length $p^s$ for any positive integer $k,s$ and all primes $p$. The paper only provided a form
for the Euclidean dual code of any cyclic code over $\mathbb{F}_{p^k}+u\mathbb{F}_{p^k}$
of length $p^s$ (see Page 9, Theorem 19 in [5]). On that basis, the paper provided formulas to
count the number of Euclidean self-dual cyclic codes and Hermitian self-dual cyclic codes over $\mathbb{F}_{p^k}+u\mathbb{F}_{p^k}$
of length $p^s$ respectively. For example, in Corollary 22(ii) of [5],
the number of Euclidean self-dual cyclic codes of length $2^s$ over $\mathbb{F}_{2^k}+u\mathbb{F}_{2^k}$ ($u^2=0$) was
$$1+2^k+2(2^k)^2\cdot\frac{(2^k)^{2^{s-2}-1}-1}{2^k-1}, \ {\rm if} \ s\geq 3.$$
But this statement requires correction. In fact, the above formula is missing a term (which proves to be quite important). In Theorem
2 of this paper, we prove the following correction: the number of Euclidean self-dual cyclic codes of length $2^s$ over $\mathbb{F}_{2^m}+u\mathbb{F}_{2^m}$ ($u^2=0$) is
$$1+2^m+2(2^m)^2\cdot\frac{(2^m)^{2^{s-2}-1}-1}{2^m-1}+(2^m)^{2^{s-2}+1}, \ {\rm when} \ s\geq 3.$$
It seems that the error was caused by a mistake in taking the geometric sum.
But the fundamental reason is
that Choosuwan et al. [5] were not given an explicit representation for
each Euclidean self-dual cyclic code of length $2^s$ over $\mathbb{F}_{2^k}+u\mathbb{F}_{2^k}$.

\par
   Recently, in \cite{s6} we  provided a new way different
from the methods used in \cite{s7}--\cite{s13} to determine the algebraic structures,
the generators and enumeration of $\lambda$-constacyclic codes over $\mathbb{F}_{p^m}+u\mathbb{F}_{p^m}$ of length
$np^s$, where $n$ is a positive integer satisfying ${\rm gcd}(p,n)=1$ and $\lambda\in \mathbb{F}_{p^m}^\times$.
Then we given an explicit representation for the dual code of  every cyclic
code and every negacyclic code. Moreover, we provided a discriminant condition for the self-duality
of each cyclic code and negacyclic code over $\mathbb{F}_{p^m}+u\mathbb{F}_{p^m}$ of length
$np^s$.

\par
  On the basis of the results obtained in [6], we can consider to give an explicit representation and enumeration for self-dual cyclic
codes and self-dual negacyclic codes over $\mathbb{F}_{p^m}+u\mathbb{F}_{p^m}$. In this paper, we focus on (Euclidean) self-dual cyclic codes over $\mathbb{F}_{2^m}+u\mathbb{F}_{2^m}$
of length $2^s$.

\vskip 2mm\par
  The present paper is organized as follows. In Section 2, we review the
known results for self-dual cyclic codes of length $2^s$ over $\mathbb{F}_{2^m}+u\mathbb{F}_{2^m}$ first. Then we prove that almost all of these self-dual cyclic codes are determined by a special kind of subsets $\Omega_l$ in the
residue class ring $\frac{\mathbb{F}_{2^m}[x]}{\langle (x+1)^l\rangle}$ for certain integers $l$, $1\leq l\leq 2^s-1$. In Section 3,
we study the calculation and representation of the set $\Omega_l$ by use of properties for Kronecker product of matrices and calculation of linear equations over $\mathbb{F}_{2^m}$. In Section 4,
we give an explicit representation and enumeration for self-dual cyclic codes over $\mathbb{F}_{2^m}+u\mathbb{F}_{2^m}$
of length $2^s$. As an application, we list precisely all distinct self-dual cyclic codes over $\mathbb{F}_{2^m}+u\mathbb{F}_{2^m}$
of length $2^s$ for $s=3,4$ in Section 5.
Section 6 concludes the paper.




\section{Preliminaries}

\noindent \par
In this section, we list the necessary notations and some known results for cyclic codes  of length $2^s$ over the ring $\mathbb{F}_{2^m}+u\mathbb{F}_{2^m}$ ($u^2=0$)
needed in the following sections.

\par Let $\mathbb{F}_{p^m}$ be a finite field of $p^m$ elements, where $p$ is a prime number, and $\lambda, \lambda_0\in \mathbb{F}_{p^m}^\times$ satisfying $\lambda_0^{p^s}=\lambda$. Then $x^{p^s}-\lambda=(x-\lambda_0)^{p^s}$. As in Section 7 of [6],
we denote
$y=x-\lambda_0$,

\begin{description}
\item{$\diamond$}
$\mathcal{A}=\frac{\mathbb{F}_{p^m}[x]}{\langle x^{p^s}-\lambda\rangle}
=\mathbb{F}_{p^m}+y\mathbb{F}_{p^m}+\ldots+y^{p^s-1}\mathbb{F}_{p^m}
\ (y^{p^s}=0).$

\item{$\diamond$}
$y^\alpha (\mathcal{A}/\langle y^\beta\rangle)
=\{\sum_{i=\alpha}^{\beta-1}a_iy^i\mid a_i\in \mathbb{F}_{p^m}, \ \alpha\leq i\leq \beta-1\}$ and $y^\alpha (\mathcal{A}/\langle y^\alpha\rangle)=\{0\}$ for all integers $0\leq \alpha<\beta\leq p^s-1$.
\end{description}

Denote $z=\widetilde{x-\lambda_0}=1-\lambda_0x$ in the following lemma.

\vskip 3mm\noindent
  {\bf Lemma 1} (cf. [6] Corollary 7.1)
\textit{Denote $y=x-\lambda_0$, $z=1-\lambda_0x$ and $q=p^m$. Then all distinct $\lambda$-constacyclic codes over $\mathbb{F}_{p^m}+u\mathbb{F}_{p^m}$ $(u^2=0)$ of length $p^s$ and their dual codes are given by
the following table}:
{\small
\begin{center}
\begin{tabular}{lll}\hline
 N     &     $\mathcal{C}$ $({\rm mod} \ x^{p^s}-\lambda)$, $\mathcal{C}^{\bot}$ $({\rm mod} \ x^{p^s}-\lambda^{-1})$ & $|\mathcal{C}|$  \\ \hline
$q^{p^s-\lceil\frac{p^s}{2}\rceil}$ & $\bullet$ $\mathcal{C}=\langle y b(x)+u\rangle$
   & $q^{p^s}$  \\
   & $\mathcal{C}^{\bot}=\langle z\overline{b(x)}+u\rangle$   & \\
   & $b(x)\in y^{\lceil \frac{p^s}{2}\rceil-1}(\mathcal{A}/\langle y^{p^s-1}\rangle)$ &  \\
$\sum_{k=1}^{p^s-1}q^{p^s-k-\lceil\frac{1}{2}(p^s-k)\rceil}$
  & $\bullet$  $\mathcal{C}=\langle y^{k+1}b(x)+u y^k\rangle$ & $q^{p^s-k}$  \\
  & $\mathcal{C}^{\bot}=\langle z\overline{b(x)}+u,z^{p^s-k}\rangle$   & \\
  & $b(x)\in y^{\lceil \frac{p^s-k}{2}\rceil-1}(\mathcal{A}/\langle y^{p^s-k-1}\rangle),$ &  \\
  & $1\leq k\leq p^s-1$  & \\
$p^s+1$ & $\bullet$  $\mathcal{C}=\langle y^k\rangle$ & $q^{2(p^s-k)}$  \\
  & $\mathcal{C}^{\bot}=\langle z^{p^s-k}\rangle$   & \\
  & $0\leq k\leq p^s$ & \\
$\sum_{t=1}^{p^s-1}q^{t-\lceil\frac{t}{2}\rceil}$ & $\bullet$  $\mathcal{C}=\langle yb(x)+u,y^{t}\rangle$
  & $q^{2p^s-t}$  \\
  & $\mathcal{C}^{\bot}=\langle z^{p^s-t+1}\overline{b(x)}+uz^{p^s-t}\rangle$   & \\
  & $b(x)\in y^{\lceil\frac{t}{2}\rceil-1}(\mathcal{A}/\langle y^{t-1}\rangle)$, $1\leq t\leq p^s-1$ & \\
$\sum_{k=1}^{p^s-2}\sum_{t=1}^{p^s-k-1}q^{t-\lceil\frac{t}{2}\rceil}$
  & $\bullet$  $\mathcal{C}=\langle y^{k+1}b(x)+u y^k,y^{k+t}\rangle$
  & $q^{2p^s-2k-t}$  \\
  & $\mathcal{C}^{\bot}=\langle z^{p^s-k-t+1}\overline{b(x)}+uz^{p^s-k-t},z^{p^s-k}\rangle$   & \\
  & $b(x)\in y^{\lceil\frac{t}{2}\rceil-1}(\mathcal{A}/\langle y^{t-1}\rangle)$,
& \\
  & $1\leq t\leq p^s-k-1$, $1\leq k\leq p^s-2$ & \\
\hline
\end{tabular}
\end{center} }

\noindent
\textit{in which $x^{-1}=\lambda x^{p^s-1}$ and $\overline{b(x)}=-\lambda x^{p^s-1}b(x^{-1})$ $({\rm mod} \
x^{p^s}-\lambda^{-1})$}.

\par
 \textit{Moreover, the number of $\lambda$-constacyclic codes over $\mathbb{F}_{p^m}+u\mathbb{F}_{p^m}$ of length $p^s$ is equal to}
$$N_{(p^m,1,p^{s})}=\left\{\begin{array}{ll}\sum_{i=0}^{2^{s-1}}(1+4i)2^{(2^{s-1}-i)m}, & {\rm if} \ p=2; \cr & \cr
                                        \sum_{i=0}^{\frac{p^s-1}{2}}(3+4i)p^{(\frac{p^s-1}{2}-i)m}, & {\rm if} \ p \ {\rm is} \ {\rm odd}. \end{array}\right.$$

\vskip 3mm\par
  Now, we consider how to determine the self-dual cyclic codes
of length $2^s$ over the ring $\mathbb{F}_{2^m}+u\mathbb{F}_{2^m}$ ($u^2=0$). We set $p=2$ and $\lambda=1$.
Then $\lambda_0=1$ and $-1=1$. This implies the following:

\begin{description}
\item{$\diamond$}
 $y=z=x+1$ and $\mathcal{A}=\frac{\mathbb{F}_{2^m}[x]}{\langle (x+1)^{2^s}\rangle}$;

\item{$\diamond$}
 $y^\alpha (\mathcal{A}/\langle y^\beta\rangle)=(x+1)^\alpha \cdot\frac{\mathbb{F}_{2^m}[x]}{\langle (x+1)^\beta\rangle}=\{\sum_{i=\alpha}^{\beta-1}a_i(x+1)^i\mid a_i\in \mathbb{F}_{2^m}, \ \alpha\leq i\leq \beta-1\}$ ($(x+1)^\beta=0$), where $0\leq \alpha<\beta\leq 2^s-1$.

\item{$\diamond$}
 $\overline{b(x)}=x^{-1}b(x^{-1})$ $({\rm mod} \
(x+1)^\beta)$ for any $b(x)\in (x+1)^\alpha \cdot\frac{\mathbb{F}_{2^m}[x]}{\langle (x+1)^\beta\rangle}$.
\end{description}

From this and by Lemma 1, we list all
$\sum_{i=0}^{2^{s-1}}(1+4i)2^{(2^{s-1}-i)m}$ cyclic codes over $\mathbb{F}_{2^m}+u\mathbb{F}_{2^m}$ of length
$2^s$ and their dual codes as follows.

\begin{description}
\item{Case I.}
  $2^{(2^s-2^{s-1})m}$ codes:

\par
  $\mathcal{C}=\langle (x+1)b(x)+u\rangle$ with $|\mathcal{C}|=2^{2^sm}$ and $\mathcal{C}^{\bot}=\langle (x+1)\cdot x^{-1}b(x^{-1})+u\rangle$,
where $b(x)\in (x+1)^{2^{s-1}-1}\cdot \frac{\mathbb{F}_{2^m}[x]}{\langle (x+1)^{2^{s}-1}\rangle}$.

\item{Case II.}
  $\sum_{k=1}^{2^s-1}2^{(2^s-k-\lceil\frac{1}{2}(2^s-k)\rceil)m}$ codes:

\par
  $\mathcal{C}=\langle (x+1)^{k+1}b(x)+u(x+1)^{k}\rangle$ with $|\mathcal{C}|=2^{(2^s-k)m}$ and $\mathcal{C}^{\bot}=\langle (x+1)\cdot x^{-1}b(x^{-1})+u, (x+1)^{2^s-k}\rangle$,
where $b(x)\in (x+1)^{\lceil\frac{2^s-k}{2}\rceil-1}\cdot \frac{\mathbb{F}_{2^m}[x]}{\langle (x+1)^{2^{s}-k-1}\rangle}$
and $1\leq k\leq 2^s-1$.

\item{Case III.}
   $2^s+1$ codes:

\par
  $\mathcal{C}=\langle (x+1)^k\rangle$ with $|\mathcal{C}|=2^{2(2^s-k)}$ and $\mathcal{C}^{\bot}=\langle (x+1)^{2^s-k}\rangle$,
where $0\leq k\leq 2^s$.

\item{Case IV.}
  $\sum_{t=1}^{2^s-1}2^{(t-\lceil\frac{t}{2}\rceil)m}$ codes:

\par
  $\mathcal{C}=\langle (x+1)b(x)+u,(x+1)^t\rangle$ with $|\mathcal{C}|=2^{(2\cdot2^s-t)m}=2^{(2^s+(2^s-t))m}$ and $\mathcal{C}^{\bot}=\langle (x+1)^{2^s-t+1}\cdot x^{-1}b(x^{-1})+u(x+1)^{2^s-t}\rangle$,
where $b(x)\in (x+1)^{\lceil\frac{t}{2}\rceil-1}\cdot \frac{\mathbb{F}_{2^m}[x]}{\langle (x+1)^{t-1}\rangle}$
and $1\leq t\leq 2^s-1$.

\item{Case V.}
  $\sum_{k=1}^{2^s-2}\sum_{t=1}^{2^s-k-1}2^{(t-\lceil\frac{t}{2}\rceil)m}$ codes:

\par
  $\mathcal{C}=\langle (x+1)^{k+1}b(x)+u (x+1)^k,(x+1)^{k+t}\rangle$ with $|\mathcal{C}|=2^{(2\cdot 2^s-2k-t)m}$
and $\mathcal{C}^{\bot}=\langle (x+1)^{2^s-k-t+1}\cdot x^{-1}b(x^{-1})+u(x+1)^{2^s-k-t},(x+1)^{2^s-k}\rangle$,
where $b(x)\in (x+1)^{\lceil\frac{t}{2}\rceil-1}\cdot \frac{\mathbb{F}_{2^m}[x]}{\langle (x+1)^{t-1}\rangle}$,
$1\leq t\leq p^s-k-1$ and $1\leq k\leq p^s-2$.
\end{description}

\vskip 3mm\par
  As $|(\mathbb{F}_{2^m}+u\mathbb{F}_{2^m})^{2^s}|=(2^{2^s m})^2$, every self-dual cyclic
code $\mathcal{C}$ over $\mathbb{F}_{2^m}+u\mathbb{F}_{2^m}$ of length $2^s$ must contain
$|\mathcal{C}|=2^{2^s m}$ codewords. From this, we deduce that there is no self-dual codes in
Cases II and IV.

\par
  Let  $\mathcal{C}=\langle (x+1)b(x)+u\rangle$ be a code in Case I. Then $\mathcal{C}=\mathcal{C}^{\bot}$ if and only if
$b(x)\in (x+1)^{2^{s-1}-1}\cdot \frac{\mathbb{F}_{2^m}[x]}{\langle (x+1)^{2^{s}-1}\rangle}$ satisfying $b(x)=x^{-1}b(x^{-1})$, i.e.,
$b(x)+x^{-1}b(x^{-1})\equiv 0$ (mod $(x+1)^{2^{s}-1}$).

\par
  Let $\mathcal{C}=\langle (x+1)^k\rangle$ be a code in Case III. Then $\mathcal{C}=\mathcal{C}^{\bot}$ if and only if
$k=2^s-k$, i.e., $k=2^{s-1}$.

\par
  Let $\mathcal{C}=\langle (x+1)^{k+1}b(x)+u (x+1)^k,(x+1)^{k+t}\rangle$ be a code in Case III.
Then $\mathcal{C}=\mathcal{C}^{\bot}$ if and only if $2\cdot 2^s-2k-t=2^s$ and $b(x)\in (x+1)^{\lceil\frac{t}{2}\rceil-1}\cdot \frac{\mathbb{F}_{2^m}[x]}{\langle (x+1)^{t-1}\rangle}$ satisfying $b(x)=x^{-1}b(x^{-1})$. The latter is equivalent to
$b(x)+x^{-1}b(x^{-1})\equiv 0$ (mod $(x+1)^{t-1}$), and the former is equivalent to $t=2^s-2k$ and $1\leq k\leq 2^{s-1}-1$.
When this condition is satisfied, we have $\lceil\frac{t}{2}\rceil=\frac{2^s-2k}{2}=2^{s-1}-k$.

\par
  In the light of the above discussion, we have the following conclusion.

\vskip 3mm\noindent
  {\bf Lemma 2 } \textit{For any integer $l$, $1\leq l\leq 2^s-1$, we denote}
\begin{itemize}
\item
$\Omega_l=\left\{b(x)\in \frac{\mathbb{F}_{2^m}[x]}{\langle (x+1)^l\rangle}
\mid b(x)+x^{-1}b(x^{-1})\equiv 0 \ ({\rm mod} \ (x+1)^l)\right\}.$
\end{itemize}
\textit{Then all distinct self-dual cyclic codes over $\mathbb{F}_{2^m}+u\mathbb{F}_{2^m}$ of length
$2^s$ are given by the following three cases}:

\begin{description}
\item{(i)}
  $\langle (x+1)^{2^{s-1}}\rangle$.

\item{(ii)}
   \textit{$\langle (x+1)b(x)+u\rangle$, where $b(x)=\sum_{i=2^{s-1}-1}^{2^s-2}b_i(x+1)^i\in \Omega_{2^s-1}$}.

\item{(iii)}
  \textit{$\langle (x+1)^{k+1}b(x)+u(x+1)^k, (x+1)^{2^s-k}\rangle$, where
$1\leq k\leq 2^{s-1}-1$ and  $b(x)=\sum_{i=2^{s-1}-k-1}^{2^s-2k-2}b_i(x+1)^i\in \Omega_{2^s-2k-1}$}.
\end{description}

\vskip 3mm\par
   In order to present all cyclic codes over $\mathbb{F}_{2^m}+u\mathbb{F}_{2^m}$ of length
$2^s$ explicitly, by Lemma 2 we need to determine the subset $\Omega_l$ of
$\frac{\mathbb{F}_{2^m}[x]}{\langle (x+1)^l\rangle}$ for $l=2^s-1$ and $l=2^s-2k-1$ where $1\leq k\leq 2^{s-1}-1$.



\section{Calculation and representation of the set $\Omega_l$} \label{}

\noindent
  In this section, we consider how to calculate effectively and represent the subset $\Omega_l$ of
$\frac{\mathbb{F}_{2^m}[x]}{\langle (x+1)^l\rangle}$ defined in Lemma 2, where $1\leq l\leq 2^s-1$.

\par
  For any matrix $A$ over $\mathbb{F}_{2^m}$, let $A^{{\rm tr}}$ be the transposition of $A$. In the rest of this paper, we adopt
the following notation
\begin{itemize}
\item
  $\mathcal{S}_l=\{B_l=(b_0,b_1,\ldots,b_{l-1})^{{\rm tr}}
\mid \sum_{i=0}^{l-1}b_i(x+1)^i\in \Omega_l, \ b_i\in \mathbb{F}_{2^m}, \ \forall i=0,1,\ldots,l-1\}$.
\end{itemize}
Then we have
$$\Omega_l=\{(1,(x+1),\ldots,(x+1)^{l-1})B_l\mid B_l\in \mathcal{S}_l\} \ {\rm and} \ |\Omega_l|=|\mathcal{S}_l|.$$
In order to present the
set $\Omega_l$ of polynomial, it is sufficiency to determine the set $\mathcal{S}_l$.

\par
   Let $A=(a_{ij})$ and $B$ be matrices over $\mathbb{F}_{2^m}$ of sizes $s\times t$ and $u\times v$ respectively.
Recall that the \textit{Kronecker product} of $A$ and $B$ is
defined by $A\otimes B=(a_{ij}B)$ which is a a matrix over $\mathbb{F}_{2^m}$ of size $su\times tv$. Then we denote
\begin{itemize}
\item
$G_2=\left(\begin{array}{cc} 1 & 0 \cr 1 & 1\end{array}\right)$;

\item
$G_{2^\lambda}=G_2\otimes G_{2^{\lambda-1}}=\left(\begin{array}{cc} G_{2^{\lambda-1}} & 0 \cr G_{2^{\lambda-1}} & G_{2^{\lambda-1}}\end{array}\right)$, $\lambda=2,3,\ldots.$
\end{itemize}

\vskip 3mm\noindent
  {\bf Theorem 1} \textit{Using the notation above, we have the following conclusions}.

\begin{description}
\item{(i)}
 \textit{We have $\mathcal{S}_1=\mathbb{F}_{2^m}$ with $|\mathcal{S}_1|=2^m$}.

\item{(ii)}
  \textit{Let $2\leq l\leq 2^s-1$ and assume $\lambda$ be the least positive integer
such that}
 $2^{\lambda-1}+1\leq l\leq 2^\lambda.$
 \textit{Let $I_{2^\lambda}$ be the
identity matrix of order $2^\lambda$, and $M_l$ be the
submatrix in the upper left corner of
$I_{2^\lambda}+G_{2^\lambda}$, i.e.,
\begin{itemize}
\item
$\left(\begin{array}{cc} M_l & 0 \cr \ast & \ast\end{array}\right)=I_{2^\lambda}+G_{2^\lambda}$,
where $M_l$ is a matrix over $\mathbb{F}_{2}$ of size $l\times l$.
\end{itemize}
Especially, we have $M_{2^\lambda}=I_{2^\lambda}+G_{2^\lambda}$.
Then $\mathcal{S}_l$ is the solution space over $\mathbb{F}_{2^m}$ of the following homogeneous linear equations}:
$$M_lY=0, \
{\rm where} \ Y=(y_0,y_1,\ldots,y_{l-1})^{{\rm tr}}.$$
\textit{Moreover, we have ${\rm dim}_{\mathbb{F}_{2^m}}(\mathcal{S}_l)=\lfloor \frac{l+1}{2}\rfloor$, i.e.,
$|\mathcal{S}_l|=2^{\lfloor \frac{l+1}{2}\rfloor m}$}.

\item{(iii)}
\textit{Let $\lambda\geq 2$ and $l=2^{\lambda-1}+\tau$, where $1\leq \tau\leq 2^{\lambda-1}$.
Then each vector $B_l\in \mathcal{S}_{l}$ can be calculated recursively by the following three steps}:

\begin{description}
\item{Step 1.}
  \textit{Choose column vector $B_{2^{\lambda-1}}=(b_0,b_1,\ldots,b_{2^{\lambda-1}-1})^{{\rm tr}}\in \mathcal{S}_{2^{\lambda-1}}$ and set $B_{2^{\lambda-1}}^{(1,\tau)}=(b_0,b_1,\ldots,b_{\tau-1})^{{\rm tr}}$}.

\item{Step 2.}
 \textit{Solving the following linear equations over $\mathbb{F}_{2^m}$:
$$M_{\tau}(y_{2^{\lambda-1}},\ldots,y_{l-1})^{{\rm tr}}=B_{2^{\lambda-1}}^{(1,\tau)}$$
where $l=2^{\lambda-1}+\tau$ and $1\leq \tau\leq 2^{\lambda-1}$, we obtain solution vectors} $$B_l^{(2^{\lambda-1}+1,l)}=(b_{2^{\lambda-1}},\ldots,b_{l-1})^{{\rm tr}}.$$

\item{Step 3.}
  \textit{Set
$B_l=\left(\begin{array}{c} B_{2^{\lambda-1}} \cr B_l^{(2^{\lambda-1}+1,l)}\end{array}\right).$}
\end{description}

  \textit{Therefore, for any integer $\nu\geq 2$ we have the following conclusions}:

\begin{itemize}
\item
 \textit{$\mathcal{S}_{2\nu-1}=\{(0,b_1,\ldots, b_{2\nu-3}, b_{2\nu-2})^{{\rm tr}} \mid b_1,b_3,\ldots,b_{2\nu-3},b_{2\nu-2}\in \mathbb{F}_{2^m}\}$, where
$b_{2h}$ is a fixed $\mathbb{F}_2$-linear combination of $b_1,b_3,\ldots, b_{2h-1}$ for all
$h=1,2,\ldots,\nu-2$}.

\item
 \textit{$\mathcal{S}_{2\nu}=\{(0,b_1,\ldots,b_{2\nu-2}, b_{2\nu-1})^{{\rm tr}}\mid b_1,b_3,\ldots,b_{2\nu-3},b_{2\nu-1}\in \mathbb{F}_{2^m}\}$, where
$b_{2h}$ is a fixed $\mathbb{F}_2$-linear combination of $b_1,b_3,\ldots, b_{2h-1}$ for all
$h=1,2,\ldots,\nu-1$}.
\end{itemize}
\end{description}

\vskip 3mm\noindent
  {\bf Proof.} Let $l$ be an integer with $1\leq l\leq 2^s-1$.
    By $x(\sum_{i=0}^{l-1}(x+1)^i)=(1+(x+1))(\sum_{i=0}^{l-1}(x+1)^i)=1+(x+1)^l\equiv 1$ (mod $(x+1)^l$), we obtain
\begin{equation}
\label{eq1}
x^{-1}\equiv\sum_{i=0}^{l-1}(x+1)^i \ ({\rm mod} \ (x+1)^l).
\end{equation}
  For simplification and clarity of the expression, in the following we denote

\begin{description}
\item{$\diamond$}
   $X_l=(1,(x+1),(x+1)^2,\ldots,(x+1)^{l-1}).$

\item{$\diamond$}
 $\xi_l=1+(x+1)+\ldots+(x+1)^{l-1}=x^{-1}$ (mod $(x+1)^l$).

\item{$\diamond$}
 $\Xi_l=(\xi_l,\xi_l^2(x+1),\ldots,\xi_l^{l}(x+1)^{l-1})=(x^{-1},x^{-2}(x+1),x^{-3}(x+1)^2$, $\ldots,x^{-l}(x+1)^{l-1})$
 (mod $(x+1)^l$).
\end{description}

\vskip 2mm\par
  (i) Let $l=1$. We have $\frac{\mathbb{F}_{2^m}[x]}{\langle x+1\rangle}=\mathbb{F}_{2^m}$,
and hence
$$\Omega_{1}=\{b(x)=a_0\in \mathbb{F}_{2^m}\mid a_0+a_0x^{-1}\equiv 0 \ ({\rm mod} \ x+1)\}=\mathbb{F}_{2^m}.$$
This implies $\mathcal{S}_1=\mathbb{F}_{2^m}$ with $|\mathcal{S}_1|=2^m$.

\vskip 2mm\par
  (ii) Let $l=2$.
 First, we prove the following statement:
\begin{equation}
\label{eq2}
\Xi_{2^\lambda}=X_{2^\lambda}G_{2^\lambda} \ ({\rm mod} \ (x+1)^{2^\lambda}), \ 1\leq \lambda\leq s-1.
\end{equation}

\par
  We use mathematical induction on $\lambda$.
   When $\lambda=1$, we have
\begin{eqnarray*}
\Xi_2&=&(1+(x+1),(1+(x+1))^2(x+1))=(1+(x+1),x+1)\\
&=&(1,x+1)\left(\begin{array}{cc} 1 & 0 \cr 1 & 1\end{array}\right)= X_2G_2 \ ({\rm mod} \ (x+1)^2).
\end{eqnarray*}
Hence Equation (\ref{eq2}) holds when $\lambda=1$.

\par
  Let $\lambda\geq 2$ and assume that
  $$\Xi_{2^{\lambda-1}}=X_{2^{\lambda-1}}G_{2^{\lambda-1}} \ ({\rm mod} \ (x+1)^{2^{\lambda-1}}).$$
Since $(x+1)^{2^{\lambda-1}}\xi_{2^{\lambda}}^{2^{\lambda-1}}
\equiv(x+1)^{2^{\lambda-1}}(\sum_{i=0}^{2^{\lambda-1}-1}(x+1)^i)^{2^{\lambda-1}}\equiv (x+1)^{2^{\lambda-1}}$
(mod $(x+1)^{2^\lambda}$), by
$(x+1)^{2^{\lambda}}\equiv 0$ (mod $(x+1)^{2^\lambda}$) and
$$\xi_{2^\lambda}=(\xi_{2^{\lambda-1}} \ {\rm mod} \ (x+1)^{2^{\lambda-1}})+(x+1)^{2^{\lambda-1}}\cdot\xi_{2^{\lambda-1}} \
 ({\rm mod} \ (x+1)^{2^\lambda}),$$
 we have
$$\xi_{2^\lambda}^i=(\xi_{2^{\lambda-1}}^i \ {\rm mod} \ (x+1)^{2^{\lambda-1}})+(x+1)^{2^{\lambda-1}}\cdot \xi_{2^{\lambda-1}}^i
\ ({\rm mod} \ (x+1)^{2^{\lambda}})$$
for all $i=1,2,\ldots, 2^{\lambda-1}$. From these, we deduce that
\begin{eqnarray*}
\Xi_{2^\lambda}&=&(\xi_{2^\lambda},\xi_{2^\lambda}^2(x+1),\ldots,\xi_{2^\lambda}^{{2^\lambda}}(x+1)^{{2^\lambda}-1})\\
 &=& \left(\xi_{2^\lambda},\xi_{2^\lambda}^2(x+1),\ldots,\xi_{2^\lambda}^{2^{\lambda-1}}(x+1)^{2^{\lambda-1}-1},\right.\\
  && \left. \xi_{2^\lambda}^{2^{\lambda-1}}(x+1)^{2^{\lambda-1}}
  \cdot\left(\xi_{2^\lambda},\xi_{2^\lambda}^2(x+1),\ldots,\xi_{2^\lambda}^{2^{\lambda-1}}(x+1)^{2^{\lambda-1}-1}\right)\right) \\
 &\equiv & \left((\Xi_{2^{\lambda-1}} \ {\rm mod} \ (x+1)^{2^{\lambda-1}})+(x+1)^{2^{\lambda-1}}(\Xi_{2^{\lambda-1}} \ {\rm mod} \ (x+1)^{2^{\lambda-1}}), \right. \\
   && \left. (x+1)^{2^{\lambda-1}}(\Xi_{2^{\lambda-1}} \ {\rm mod} \ (x+1)^{2^{\lambda-1}})\right)\\
 &\equiv & \left(X_{2^{\lambda-1}}G_{2^{\lambda-1}}+(x+1)^{2^{\lambda-1}}X_{2^{\lambda-1}}G_{2^{\lambda-1}},
    (x+1)^{2^{\lambda-1}}X_{2^{\lambda-1}}G_{2^{\lambda-1}}\right)\\
 &=& (X_{2^{\lambda-1}},(x+1)^{2^{\lambda-1}}X_{2^{\lambda-1}})\left(\begin{array}{cc} G_{2^{\lambda-1}} & 0
 \cr G_{2^{\lambda-1}} & G_{2^{\lambda-1}}\end{array}\right) \\
 &=& X_{2^\lambda}G_{2^\lambda} \ ({\rm mod} \ (x+1)^{2^\lambda}).
\end{eqnarray*}
Hence Equation (\ref{eq2}) holds for all $\lambda$.

\vskip 2mm\par
   Then we determine the set $\mathcal{S}_l$, where $2^{\lambda-1}+1\leq l\leq 2^\lambda$ and $2\leq \lambda\leq s-1$.
  Let $b(x)\in \frac{\mathbb{F}_{2^m}[x]}{\langle (x+1)^l\rangle}$. Then $b(x)$ has a unique $(x+1)$-expansion:
\begin{equation}
\label{eq3}
b(x)=\sum_{i=0}^{l-1}b_i(x+1)^i=X_lB_l, \ B_l=(b_0,b_1,\ldots,b_{l-1})^{{\rm tr}}
\ {\rm with} \ b_{i}\in \mathbb{F}_{2^m}.
\end{equation}
From this and by Equation (\ref{eq1}), we deduce that
$$x^{-1}b(x^{-1})=x^{-1}\sum_{i=0}^{l-1}b_i(x^{-1}+1)^i=\sum_{i=0}^{l-1}b_ix^{-(i+1)}(x+1)^i
=\sum_{i=0}^{l-1}b_i\xi_l^{i+1}(x+1)^i,$$
i.e., $x^{-1}b(x^{-1})=\Xi_lB_l$. From these, we deduce that
\begin{eqnarray*}
b(x)+x^{-1}b(x^{-1})
 &\equiv& X_lB_l+\Xi_lB_l=(X_{2^\lambda}+\Xi_{2^\lambda})\left(\begin{array}{c}B_l \cr 0\end{array}\right)\\
 &\equiv & X_{2^\lambda}(I_{2^\lambda}+G_{2^\lambda})\left(\begin{array}{c}B_l \cr 0\end{array}\right)\\
&\equiv& (X_l,0)\left(\begin{array}{cc} M_l & 0 \cr \ast & \ast\end{array}\right)\left(\begin{array}{c}B_l \cr 0\end{array}\right)\\
&\equiv& X_l(M_lB_l)  \ ({\rm mod} \ (x+1)^l).
\end{eqnarray*}
Then by the definition of $\Omega_{l}$ in Lemma 2 and Equation (\ref{eq3}), we see that
$b(x)\in \Omega_{l}$ if and only if $M_lB_l=0$. The latter is equivalent to that
$B_l$ is a solution vector of $M_lY=0$. Hence
$\mathcal{S}_l$ is the solution space over $\mathbb{F}_{2^m}$ of the homogeneous linear equations
$M_lY=0$, by the definition of $\mathcal{S}_l$.

\par
   By $G_{2^\lambda}=\left(\begin{array}{cc} G_{2^{\lambda-1}} & 0 \cr G_{2^{\lambda-1}} & G_{2^{\lambda-1}}\end{array}\right)$,
$G_2=\left(\begin{array}{cc} 1 & 0 \cr 1 & 1 \end{array}\right)$ and $\left(\begin{array}{cc} M_l & 0 \cr \ast & \ast\end{array}\right)
=I_{2^\lambda}+G_{2^\lambda}$, we see that
$M_l$ is a strictly lower triangular matrix and ${\rm rank}(M_l)=\lceil \frac{l-1}{2}\rceil$.
Since $\mathcal{S}_l$ is the solution space of homogeneous linear equations $M_lY=0$
over $\mathbb{F}_{2^m}$, we have that ${\rm dim}_{\mathbb{F}_{2^m}}(\mathcal{S}_l)=l-\lceil \frac{l-1}{2}\rceil
=1+\lfloor\frac{l-1}{2}\rfloor=\lfloor\frac{l+1}{2}\rfloor$  (For similar results, see [18, Proposition 3.3]). This implies
$|\Omega_{l}|=|\mathcal{S}_l|=2^{\lfloor\frac{l+1}{2}\rfloor m}$.

\par
 (iii) Let $\lambda\geq 2$ and $l=2^{\lambda-1}+\tau$, where $1\leq \tau\leq 2^{\lambda-1}$. By (ii) we know
that $M_{2^\lambda}=I_{2^\lambda}+G_{2^\lambda}=\left(\begin{array}{cc}M_{2^{\lambda-1}} & 0
\cr  I_{2^{\lambda-1}}+M_{2^{\lambda-1}} & M_{2^{\lambda-1}}\end{array}\right).$ This implies
$$M_l=\left(\begin{array}{cc}M_{2^{\lambda-1}} & 0
\cr  (I_\tau,0)+(M_\tau,0) & M_{\tau}\end{array}\right).$$
For each $B_l\in\mathcal{S}_{l}$, we write $B_l=\left(\begin{array}{c} B_l^{(1,2^{\lambda-1})} \cr B_l^{(2^{\lambda-1}+1,l)}\end{array}\right)$
where
$$B_l^{(1,2^{\lambda-1})}=\left(\begin{array}{c} b_0\cr b_1\cr \vdots \cr b_{2^{\lambda-1}-1}\end{array}\right),
 \ B_l^{(2^{\lambda-1}+1,l)}=\left(\begin{array}{c} b_{2^{\lambda-1}}\cr \vdots \cr b_{l-1}\end{array}\right)
\ {\rm and} \ B_l^{(1,\tau)}=\left(\begin{array}{c} b_0\cr b_1\cr \vdots \cr b_{\tau-1}\end{array}\right).$$
Then by (ii), we have $M_lB_l=0$, i.e.,
$$\left\{\begin{array}{l} M_{2^{\lambda-1}}B_l^{(1,2^{\lambda-1})}= 0; \cr
B_l^{(1,\tau)}+M_{\tau}B_l^{(1,\tau)}+ M_{\tau}B_l^{(2^{\lambda-1}+1,l)}=0\end{array}\right.$$
From the first equation, we deduce that $B_l^{(1,2^{\lambda-1})}\in \mathcal{S}_{2^{\lambda-1}}$. This implies
$M_{\tau}B_l^{(1,\tau)}=0$ by (ii) and $1\leq \tau\leq 2^{\lambda-1}$. Then by the second
equation, it follows that
\begin{equation}
\label{eq4}
M_{\tau}B_l^{(2^{\lambda-1}+1,l)}=B_l^{(1,\tau)}.
\end{equation}
Now, denote $\mathcal{S}_{2^{\lambda-1}}^\prime=\{B_l^{(1,2^{\lambda-1})}\mid B_l\in \mathcal{S}_l\}$. Then we
have $\mathcal{S}_{2^{\lambda-1}}^\prime\subseteq \mathcal{S}_{2^{\lambda-1}}$.

\par
  As the number of solutions
of the linear equations (\ref{eq4}) is equal to $|\mathcal{S}_\tau|=2^{\lfloor\frac{\tau+1}{2}\rfloor m}$ by linear algebra theory and (ii), it follows
that
$$2^{\lfloor\frac{l+1}{2}\rfloor m}=|\mathcal{S}_l|=|\mathcal{S}_{2^{\lambda-1}}^\prime||\mathcal{S}_\tau|=2^{\lfloor\frac{\tau+1}{2}\rfloor m}|\mathcal{S}_{2^{\lambda-1}}^\prime|.$$
From this and by $|\mathcal{S}_l|=2^{\lfloor\frac{l+1}{2}\rfloor m}$ and $l=2^{\lambda-1}+\tau$, we deduce that
$$|\mathcal{S}_{2^{\lambda-1}}^\prime|=2^{(\lfloor\frac{2^{\lambda-1}+\tau+1}{2}\rfloor-\lfloor\frac{\tau+1}{2}\rfloor)m}=
2^{2^{\lambda-2}m}=2^{\lfloor\frac{2^{\lambda-1}+1}{2}\rfloor m}=|\mathcal{S}_{2^{\lambda-1}}|.$$
Therefore, we have that $\mathcal{S}_{2^{\lambda-1}}^\prime=\mathcal{S}_{2^{\lambda-1}}$. Hence the linear
equations (\ref{eq4}) are equivalent to
$$M_{\tau}B_l^{(2^{\lambda-1}+1,l)}=B_{2^{\lambda-1}}^{(1,\tau)},$$
where $B_{2^{\lambda-1}}=(b_0,b_1,\ldots,b_{2^{\lambda-1}-1})^{{\rm tr}}\in \mathcal{S}_{2^{\lambda-1}}$, i.e.,
$$\mathcal{S}_{2^{\lambda-1}}=\{B_l^{(1,2^{\lambda-1})}\mid B_l\in \mathcal{S}_l\} \
({\rm for} \ {\rm all} \ l=2^{\lambda-1}+\tau \ {\rm and} \ 1\leq \tau\leq 2^{\lambda-1}),$$
$B_{2^{\lambda-1}}^{(1,\tau)}=(b_0,b_1,\ldots,b_{\tau-1})^{{\rm tr}}$
and the vector $B_l^{(2^{\lambda-1}+1,l)}$ satisfies $\left(\begin{array}{c} B_{2^{\lambda-1}} \cr B_l^{(2^{\lambda-1}+1,l)}\end{array}\right)$ $\in\mathcal{S}_l$.

\par
  As stated above, we conclude that
$B_l=B_{2^{\lambda-1}+\tau}=\left(\begin{array}{c}B_{2^{\lambda-1}}\cr B_l^{(2^{\lambda-1}+1,l)}\end{array}\right)$
where $B_{2^{\lambda-1}}\in \mathcal{S}_{2^{\lambda-1}}$ and $B_l^{(2^{\lambda-1}+1,l)}$ is a solution
of the linear equations $M_{\tau}Y=B_{2^{\lambda-1}}^{(1,\tau)}$, for all
$1\leq \tau\leq 2^{\lambda-1}$.

\par
  Finally, the conclusion for the structure and properties of the $\mathbb{F}_{2^m}$-linear space $\mathcal{S}_l$ can be
  proved by the mathematical induction on the integer $\lambda\geq 2$. Here, we omit the specific processes.
\hfill $\Box$



\section{Representation and enumeration for self-dual cyclic codes} \label{}

\noindent
In this section, we determine all self-dual cyclic codes of length $2^s$ over $\mathbb{F}_{2^m}+u\mathbb{F}_{2^m}$ ($u^2=0$).

\vskip 3mm\noindent
 {\bf Theorem 2}
   \textit{For any integers $l$ and $\delta$, $0\leq \delta<l\leq 2^s-1$, denote
\begin{itemize}
\item
$\mathcal{S}_l^{[\delta]}=\{B_l^{[\delta]}=(b_\delta\, \ldots, b_{l-1})^{{\rm tr}}\mid
(0,\ldots,0,b_\delta\, \ldots, b_{l-1})^{{\rm tr}} \in\mathcal{S}_l\}.$
\end{itemize}
Then}

\begin{description}
\item{(i)}
  \textit{We have the following conclusions}:

\begin{description}
\item{(i-1)}
  \textit{When $s=1$, we have $\mathcal{S}_{2^s-1}^{[2^{s-1}-1]}=\mathcal{S}_{1}^{[0]}=\mathbb{F}_{2^m}$}.

\item{(i-2)}
  \textit{When $s=2$, we have $\mathcal{S}_{2^s-1}^{[2^{s-1}-1]}=\mathcal{S}_{3}^{[1]}=\{(b_1,b_2)^{{\rm tr}}\mid b_1,b_2\in \mathbb{F}_{2^m}\}$}.

\item{(i-3)}
 \textit{When $s\geq 3$, we have}

  \textit{$\mathcal{S}_{2^s-1}^{[2^{s-1}-1]}=\{(b_{2^{s-1}-1},b_{2^{s-1}},\ldots, b_{2^s-3}, b_{2^s-2})^{{\rm tr}}\mid b_{2^{s-2}+2i-1}, b_{2^s-2}\in \mathbb{F}_{2^m}$, $i=0,1,\ldots,2^{s-2}-1\}$}
   \textit{in which
 $b_{2^{s-1}+2i}$ is a fixed $\mathbb{F}_{2^m}$-linear combination of $b_{2^{s-1}+2z-1}$, $0\leq z\leq i$, for all $i=0,1,\ldots,2^{s-2}-2$. Hence}
 $$|\mathcal{S}_{2^s-1}^{[2^{s-1}-1]}|=(2^m)^{2^{s-2}+1}.$$

\vskip 2mm\par
  $\diamondsuit$ \textit{When $k=2^{s-1}-1$, $\mathcal{S}_{2^s-2k-1}^{[2^{s-1}-k-1]}=\mathcal{S}_{1}^{[0]}=\mathbb{F}_{2^m}$}.

\vskip 2mm\par
  $\diamondsuit$ \textit{Let $1\leq k\leq 2^{s-1}-2$. We have the following two cases}.

\begin{description}
\item{$\diamond$}
  \textit{If $k=2^{s-1}-2h$ with $1\leq h\leq 2^{s-2}-1$, we have}

  \textit{$\mathcal{S}_{2^s-2k-1}^{[2^{s-1}-k-1]}=\mathcal{S}_{4h-1}^{[2h-1]}
=\{(b_{2h-1},b_{2h},b_{2h+1},\ldots,b_{4h-3}$, $b_{4h-2})^{{\rm tr}}\mid b_{2h+2i-1}\in \mathbb{F}_{2^m}$,
 $i=0,1,\ldots,h-1, \ {\rm and} \ b_{4h-2}\in \mathbb{F}_{2^m}\}$}
 \textit{in which
 $b_{2h+2i}$ is a fixed $\mathbb{F}_{2^m}$-linear combination of $b_{2h+2z-1}$, $0\leq z\leq i$, for all $i=0,1,\ldots,h-2$. Hence}
 $$|\mathcal{S}_{2^s-2k-1}^{[2^{s-1}-k-1]}|=|\mathcal{S}_{4h-1}^{[2h-1]}|=(2^m)^{h+1}.$$

\item{$\diamond$}
 \textit{If $k=2^{s-1}-2h-1$ with $1\leq h\leq 2^{s-2}-1$, we have}

 \textit{$\mathcal{S}_{2^s-2k-1}^{[2^{s-1}-k-1]}=\mathcal{S}_{4h+1}^{[2h]}
=\{(0,b_{2h+1},\ldots,b_{4h-1}$, $b_{4h})^{{\rm tr}}\mid b_{2h+1+2i}\in \mathbb{F}_{2^m}, \
 i=0,1,\ldots,h-1, \ {\rm and} \ b_{4h}\in \mathbb{F}_{2^m}\}$}
 \textit{in which
 $b_{2h+2+2i}$ is a fixed $\mathbb{F}_{2^m}$-linear combination of $b_{2h+2+2z-1}$, $0\leq z\leq i$, for all $i=0,1,\ldots, h-2$. Hence}
 $$|\mathcal{S}_{2^s-2k-1}^{[2^{s-1}-k-1]}|=|\mathcal{S}_{4h+1}^{[4h]}|=(2^m)^{h+1}.$$
\end{description}
\end{description}

\item{(ii)}
   \textit{All distinct
self-dual cyclic codes
over $\mathbb{F}_{2^m}+u\mathbb{F}_{2^m}$ of length $2$ are given by the following}:

\vskip 2mm\par
   $\langle x+1\rangle$;

\vskip 2mm\par
   \textit{$\langle (x+1)b+u\rangle$ where $b\in \mathbb{F}_{2^m}$}.

\vskip 2mm\noindent
  \textit{Therefore, the number of self-dual cyclic codes
over $\mathbb{F}_{2^m}+u\mathbb{F}_{2^m}$ of length $2$ is $N_1=1+2^m$}.

\item{(iii)}
   \textit{All distinct
self-dual cyclic codes
over $\mathbb{F}_{2^m}+u\mathbb{F}_{2^m}$ of length $4$ are given by the following}:

\vskip 2mm\par
   $\langle (x+1)^2\rangle$;

\vskip 2mm\par
   \textit{$\langle (x+1)b(x)+u\rangle$, where $b(x)=b_1(x+1)+b_2(x+1)^2$ with $b_1,b_2\in \mathbb{F}_{2^m}$};

\vskip 2mm\par
   \textit{$\langle (x+1)^2b+u(x+1),(x+1)^3\rangle$ where $b\in \mathbb{F}_{2^m}$}.

\vskip 2mm\noindent
  \textit{Therefore, the number of self-dual cyclic codes
over $\mathbb{F}_{2^m}+u\mathbb{F}_{2^m}$ of length $4$ is $N_2=1+2^m+(2^m)^2$}.

\item{(iv)}
   \textit{Let $s\geq 3$. All distinct
self-dual cyclic codes
over $\mathbb{F}_{2^m}+u\mathbb{F}_{2^m}$ of length $2^s$ are given by the following}:

\begin{description}
\item{$\diamond$}
  $\langle (x+1)^{2^{s-1}}\rangle$.

\item{$\diamond$}
  \textit{$\langle (x+1)b(x)+u\rangle$, where $b(x)=\sum_{i=2^{s-1}-1}^{2^s-2}b_i(x+1)^i$
with $(b_{2^{s-1}-1},b_{2^{s-1}}$, $\ldots,b_{2^s-2})^{{\rm tr}}\in \mathcal{S}_{2^s-1}^{[2^{s-1}-1]}$}.

\item{$\diamond$}
  \textit{$\langle (x+1)^{2^{s-1}}b+u(x+1)^{2^{s-1}-1},(x+1)^{2^{s-1}+1}\rangle$ where $b\in \mathbb{F}_{2^m}$}.

\item{$\diamond$}
  \textit{$\langle (x+1)^{2^{s-1}-2h+1}b(x)+u(x+1)^{2^{s-1}-2h},(x+1)^{2^{s-1}+2h}\rangle$, where
  $b(x)=\sum_{i=2h-1}^{4h-2}b_i(x+1)^i$,
  $(b_{2h-1},b_{2h},b_{2h+1},\ldots, b_{4h-3},b_{4h-2})^{{\rm tr}}\in \mathcal{S}_{4h-1}^{[2h-1]}$
  and $1\leq h\leq 2^{s-2}-1$}.

\item{$\diamond$}
  \textit{$\langle (x+1)^{2^{s-1}-2h}b(x)+u(x+1)^{2^{s-1}-2h-1},(x+1)^{2^{s-1}+2h+1}\rangle$, where
  $b(x)=\sum_{i=2h+1}^{4h}b_i(x+1)^i$,
  $(0,b_{2h+1},b_{2h+2},\ldots, b_{4h-1},b_{4h})^{{\rm tr}}\in \mathcal{S}_{4h+1}^{[2h]}$
  and $1\leq h\leq 2^{s-2}-1$}.
\end{description}

  \textit{Therefore, the number of self-dual cyclic codes
over $\mathbb{F}_{2^m}+u\mathbb{F}_{2^m}$ of length $2^s$ is}
$$N_s=1+2^m+2(2^m)^2\cdot\frac{(2^m)^{2^{s-2}-1}-1}{2^m-1}+(2^m)^{2^{s-2}+1}, \ s\geq 3.$$
\end{description}

\vskip 3mm\noindent
  {\bf Proof.}
  (\dag) For each $B_l=(b_0,b_1,\ldots,b_{l-1})^{{\rm tr}}\in \mathcal{S}_l$, by Theorem 1(iii) we know that
$b_{2h}$ is a fixed $\mathbb{F}_{2^m}$-linear combination of $b_{2i-1}$, $i=1,\ldots,h$,
for all $1\leq h\leq \lceil \frac{l-1}{2}\rceil$. By the definition of
$B_l^{[\delta]}=(b_\delta,\ldots,b_{l-1})^{{\rm tr}}$, we have that $b_{2i-1}=0$ for all $i=1,\ldots,\lceil \frac{\delta-1}{2}\rceil$.
From this, we deduce the following conclusions:

\par
  $\triangleright$ When $l=1$ and $\delta=0$, we have $B_1^{[0]}=b_0\in \mathbb{F}_{2^m}$.

\par
  $\triangleright$ When $l\geq 2$ and $\delta$ is odd, we have that
$b_{2h}$ is a fixed $\mathbb{F}_{2^m}$-linear combination of $b_{2i-1}$, $i=\frac{\delta+1}{2},\ldots,h$,
for all $\frac{\delta+1}{2}\leq h\leq \lceil \frac{l-1}{2}\rceil$.

\par
  $\triangleright$ When $l\geq 3$ and $\delta$ is even, we have that $b_\delta=0$ and $b_{2h}$ is a fixed $\mathbb{F}_{2^m}$-linear combination of $b_{2i-1}$, $i=\frac{\delta}{2}+1,\ldots,h$,
for all $\delta+1\leq h\leq \lceil \frac{l-1}{2}\rceil$.

\par
  From these, one can easily verify that all conclusions in (i) are hold. Here, we omit these details.

\vskip 3mm\par
  ($\ddag$) By Lemma 2, all distinct
self-dual codes over $\mathbb{F}_{2^m}+u\mathbb{F}_{2^m}$ of length $2^s$ are given by one of the following three cases:

\par
  {\bf Case 1}. $C=\langle (x+1)^{2^{s-1}}\rangle$.

\par
  {\bf Case 2}. $\langle (x+1)b(x)+u\rangle$, where $b(x)=\sum_{i=2^{s-1}-1}^{2^s-2}b_i(x+1)^i\in \Omega_{2^s-1}$.

\par
   Let $b(x)=\sum_{i=2^{s-1}-1}^{2^s-2}b_i(x+1)^i\in \Omega_{2^s-1}$. By the definition of $\mathcal{S}_{2^s-1}$
in the beginning of section 3, we see that
$B_{2^s-1}=(0,\ldots,0,b_{2^{s-1}-1},\ldots, b_{2^s-2})^{{\rm tr}}\in \mathcal{S}_{2^s-1}$, which is equivalent
to that $(b_{2^{s-1}-1},\ldots, b_{2^s-2})^{{\rm tr}}\in \mathcal{S}_{2^s-1}^{[2^{s-1}-1]}$.

\par
  {\bf Case 3}. $\langle (x+1)^{k+1}b(x)+u(x+1)^k, (x+1)^{2^s-k}\rangle$, where
$1\leq k\leq 2^{s-1}-1$ and $b(x)=\sum_{i=2^{s-1}-k-1}^{2^s-2k-2}b_i(x+1)^i\in \Omega_{2^s-2k-1}$.

\par
  Let $b(x)=\sum_{i=2^{s-1}-k-1}^{2^s-2k-2}b_i(x+1)^i\in \Omega_{2^s-2k-1}$. Then by the definition of $\mathcal{S}_{2^s-2k-1}$, we
have $B_{2^s-2k-1}=(0,\ldots,0,b_{2^{s-1}-k-1},\ldots, b_{2^s-2k-2})^{{\rm tr}}\in \mathcal{S}_{2^s-2k-1}$, which is equivalent
to that $(b_{2^{s-1}-k-1},\ldots, b_{2^s-2k-2})^{{\rm tr}}\in \mathcal{S}_{2^s-2k-1}^{[2^{s-1}-k-1]}$.

\vskip 2mm\par
  (ii) Let $s=1$. Then there is no integer $k$ satisfying $1\leq k\leq 2^{s-1}-1=0$. Hence there is no codes in Case 3.
From Cases 1 and 2, by $\mathcal{S}_{2^s-1}^{[2^{s-1}]}=\mathcal{S}_1^{[0]}=\mathbb{F}_{2^m}$ we deduce that
all $1+2^m$ distinct self-dual cyclic codes over $\mathbb{F}_{2^m}+u\mathbb{F}_{2^m}$ of length $2$ are
given by: $\langle x+1\rangle$; $\langle (x+1)b+u\rangle$ where $b\in \mathbb{F}_{2^m}$.

\vskip 2mm\par
  (iii) Let $s=2$. By $1\leq k\leq 2^{s-1}-1=1$, we have $k=1$.
Then from the expression for codes in Cases 1--3 and by
$$\mathcal{S}_{2^s-1}^{[2^{s-1}-1]}=\mathcal{S}_{3}^{[1]}=\{(b_1,b_2)^{{\rm tr}}\mid b_1,b_2\in \mathbb{F}_{2^m}\} \ {\rm and}
\ \mathcal{S}_{2^s-2k-1}^{[2^{s-1}-k-1]}=\mathcal{S}_{1}^{\left[0\right]}=\mathbb{F}_{2^m},$$
we deduce that
all distinct $1+2^m+(2^m)^2$ self-dual cyclic codes
over $\mathbb{F}_{2^m}+u\mathbb{F}_{2^m}$ of length $2^2$ are given by:
   $\langle (x+1)^2\rangle$;
   $\langle (x+1)b(x)+u\rangle$ where $b(x)=b_1(x+1)+b_2(x+1)^2$ with $b_1,b_2\in \mathbb{F}_{2^m}$;
   $\langle (x+1)^2b+u(x+1), (x+1)^3\rangle$ where $b\in \mathbb{F}_{2^m}$.

\vskip 2mm\par
  (iv) Let $s\geq 3$. Then the codes in Case 2 are given by:
   $$\langle (x+1)b(x)+u\rangle,$$
where $b(x)=\sum_{i=2^{s-1}-1}^{2^s-2}b_i(x+1)^i$
with $(b_{2^{s-1}-1},\ldots, b_{2^s-2})^{{\rm tr}}\in \mathcal{S}_{2^s-1}^{[2^{s-1}-1]}$. Hence
the number of codes in Case 1 is $|\mathcal{S}_{2^s-1}^{[2^{s-1}-1]}|=(2^m)^{2^{s-2}+1}$.

\par
  In the following, we determine the codes in Case 3. In this case, we have $1\leq k\leq 2^{s-1}-1$.
Then we have
the following two subcases of Case 3:

\par
  {\bf Subcase 3-1}. $k=2^{s-1}-1$. By $\mathcal{S}_{2^s-2k-1}^{[2^{s-1}-k-1]}=\mathcal{S}_{1}^{[0]}=\mathbb{F}_{2^m}$,
we see that
there are $2^m$ codes in this subcase given by:
$$\langle (x+1)^{2^{s-1}}b+u(x+1)^{2^{s-1}-1}, (x+1)^{2^{s-1}+1}\rangle \
{\rm where} \ b\in \mathbb{F}_{2^m}.$$

\par
  {\bf Subcase 3-2}. $1\leq k=2^{s-1}-2$. We set $\nu=2^{s-1}-k$. Then $k=2^{s-1}-\nu$, where $2\leq\nu\leq 2^{s-1}-1$,
and we have the following two cases:

\par
   $\diamond$ When $\nu=2h$ with $h\geq 1$, we have $k=2^{s-1}-2h$ where $1\leq h\leq 2^{s-2}-1$.
Hence $\mathcal{S}_{2^s-2k-1}^{[2^{s-1}-k-1]}=\mathcal{S}_{4h-1}^{[2h-1]}$. In this case, there are
$|\mathcal{S}_{4h-1}^{[2h-1]}|=(2^m)^{h+1}$ codes:
\begin{eqnarray*}
 &&\langle (x+1)^{(2^{s-1}-2h)+1}b(x)+u(x+1)^{2^{s-1}-2h},(x+1)^{2^s-(2^{s-1}-2h)}\rangle\\
&=&\langle (x+1)^{2^{s-1}-2h+1}b(x)+u(x+1)^{2^{s-1}-2h},(x+1)^{2^{s-1}+2h}\rangle,
\end{eqnarray*}
where
  $b(x)=b_{2h-1}(x+1)^{2h-1}+b_{2h}(x+1)^{2h}+\ldots+b_{4h-3}(x+1)^{4h-3}+ b_{4h-2}(x+1)^{4h-2}$,
with $(b_{2h-1},b_{2h},\ldots, b_{4h-3},b_{4h-2})^{{\rm tr}}\in \mathcal{S}_{4h-1}^{[2h-1]}$.

\par
  $\diamond$ When $\nu=2h+1$ with $h\geq 1$, we have $k=2^{s-1}-2h-1$ where $1\leq h\leq 2^{s-2}-1$.
Hence $\mathcal{S}_{2^s-2k-1}^{[2^{s-1}-k-1]}=\mathcal{S}_{4h+1}^{[2h]}$. In this case, there are
$|\mathcal{S}_{4h+1}^{[2h]}|=(2^m)^{h+1}$ codes given by:
\begin{eqnarray*}
 &&\langle (x+1)^{(2^{s-1}-2h-1)+1}b(x)+u(x+1)^{2^{s-1}-2h-1},(x+1)^{2^s-(2^{s-1}-2h-1)}\rangle\\
&=&\langle (x+1)^{2^{s-1}-2h}b(x)+u(x+1)^{2^{s-1}-2h-1},(x+1)^{2^{s-1}+2h+1}\rangle,
\end{eqnarray*}
where
  $b(x)=b_{2h+1}(x+1)^{2h+1}+b_{2h+2}(x+1)^{2h+1}+\ldots+b_{4h-1}(x+1)^{4h-1}+ b_{4h}(x+1)^{4h}$
and $(0,b_{2h+1},b_{2h+2},\ldots, b_{4h-1},b_{4h})^{{\rm tr}}\in \mathcal{S}_{4h+1}^{[2h]}$.

\par
  As stated above, we see that
the number $N_s$ of self-dual cyclic codes over $\mathbb{F}_{2^m}+u\mathbb{F}_{2^m}$ of length $2^s$ is equal to
$$N_s=1+2^m+2\sum_{h=1}^{2^{s-2}-1}(2^m)^{h+1}+(2^m)^{2^{s-2}+1},$$
where $\sum_{h=1}^{2^{s-2}-1}(2^m)^{h+1}=(2^m)^2\sum_{i=0}^{2^{s-2}-2}(2^m)^{i-1}
=(2^m)^2\cdot\frac{(2^m)^{2^{s-2}-1}-1}{2^m-1}$.
\hfill $\Box$

\section{Self-dual cyclic codes over $\mathbb{F}_{2^m}+u\mathbb{F}_{2^m}$ of length $2^3$ and $2^4$}
\noindent
  In this section, we show how to list precisely all distinct
self-dual cyclic codes over $\mathbb{F}_{2^m}+u\mathbb{F}_{2^m}$ of length $2^s$ for specific positive
integer $s$.

\par
  Here, we take cases of $s=3,4$ as examples. We first calculate sets $\mathcal{S}_l$ ($1\leq l\leq 16$) by Theorem 1(iii) and we
put the calculation process and results in the appendix of this paper.
Then by use of Theorem 2, we provide an explicit representation for self-dual cyclic codes
over $\mathbb{F}_{2^m}+u\mathbb{F}_{2^m}$ of length $2^3$ and $2^4$, respectively.

\vskip 2mm\par
  $\diamondsuit$ Let $s=3$. By Appendix of this paper, we have

\vskip 2mm\par
  $\mathcal{S}_{2^s-1}^{[2^{s-1}-1]}=\mathcal{S}_{7}^{[3]}=\{(b_3,0,b_5,b_6)^{{\rm tr}}\mid b_3,b_5,b_6\in \mathbb{F}_{2^m}\}$.

\par
  $\mathcal{S}_{1}^{[0]}=\mathbb{F}_{2^m}$, $\mathcal{S}_{4-1}^{\left[2-1\right]}=\mathcal{S}_{3}^{\left[1\right]}
  =\{(b_1,b_2)^{{\rm tr}}\mid b_1,b_2\in \mathbb{F}_{2^m}\}$,

\par
  $\mathcal{S}_{4+1}^{\left[2\right]}
  =\mathcal{S}_{5}^{\left[2\right]}
  =\{(0,b_3,b_4)^{{\rm tr}}\mid b_3,b_4\in \mathbb{F}_{2^m}\}$.

\vskip 2mm\noindent
From these and by Theorem 2(iv), we list all distinct
$1+2^m+2(2^m)^2+(2^m)^3$ self-dual cyclic codes
over $\mathbb{F}_{2^m}+u\mathbb{F}_{2^m}$ of length $2^3$ as follows:

\begin{description}
\item{$\diamond$}
   $\langle (x+1)^4\rangle$;

\item{$\diamond$}
    $\langle(x+1)b(x)+u\rangle$,
    where $b(x)=b_3(x+1)^3+b_5(x+1)^5+b_6(x+1)^6$ with $b_3,b_5,b_6\in \mathbb{F}_{2^m}$;

\item{$\diamond$}
   $\langle (x+1)^4b+u(x+1)^3, (x+1)^5\rangle$ where $b\in \mathbb{F}_{2^m}$;

\item{$\diamond$}
  $\langle (x+1)^3b(x)+u(x+1)^2, (x+1)^6\rangle$, where $b(x)=b_1(x+1)+b_2(x+1)^2$ with $b_1, b_2\in \mathbb{F}_{2^m}$;

\item{$\diamond$}
  $\langle (x+1)^2b(x)+u(x+1), (x+1)^7\rangle$, where $b(x)=b_3(x+1)^3+b_4(x+1)^4$ with $b_3, b_4\in \mathbb{F}_{2^m}$.
\end{description}

  $\diamondsuit$ Let $s=4$. By Appendix of this paper, we have

\vskip 2mm\par
  $\mathcal{S}_{2^s-1}^{[2^{s-1}-1]}=\mathcal{S}_{15}^{[7]}=\{(b_7,0,b_9,b_9,b_{11},b_9,b_{13},b_{14})^{{\rm tr}}\mid b_7,b_9,b_{11},b_{13},b_{14}\in \mathbb{F}_{2^m}\}$.

\par
  $\mathcal{S}_{8-1}^{\left[4-1\right]}
  =\mathcal{S}_{7}^{\left[3\right]}
  =\{(b_3,0,b_5,b_6)^{{\rm tr}}\mid b_3,b_5,b_6\in \mathbb{F}_{2^m}\}$,

\par
  $\mathcal{S}_{12-1}^{\left[6-1\right]}
  =\mathcal{S}_{11}^{\left[5\right]}
  =\{(b_5,b_5,b_7,0,b_9,b_{10})^{{\rm tr}}\mid b_5,b_7,b_9,b_{10}\in \mathbb{F}_{2^m}\}$;

\par
  $\mathcal{S}_{8+1}^{\left[4\right]}
  =\mathcal{S}_{9}^{\left[4\right]}
  =\{(0,b_5,b_5,b_7,b_8)^{{\rm tr}}\mid b_5,b_7,b_8\in \mathbb{F}_{2^m}\}$,

\par
  $\mathcal{S}_{12+1}^{\left[6\right]}
  =\mathcal{S}_{13}^{\left[6\right]}
  =\{(0,b_7,0,b_9,b_9,b_{11},b_{12})^{{\rm tr}}\mid b_7,b_9,b_{11},b_{12}\in \mathbb{F}_{2^m}\}$.

\vskip 2mm\noindent
From these and by Theorem 2(iv), we list all distinct
$$1+2^m+2(2^m)^2(1+2^m+(2^m)^2)+(2^m)^5$$
self-dual cyclic codes
over $\mathbb{F}_{2^m}+u\mathbb{F}_{2^m}$ of length $2^4$ as follows:

\begin{description}
\item{$\diamond$}
  $\langle (x+1)^8\rangle$;

\item{$\diamond$}
  $\langle(x+1)b(x)+u\rangle$, where $b(x)=b_7(x+1)^7+b_9(x+1)^9+b_9(x+1)^{10}+b_{11}(x+1)^{11}+b_9(x+1)^{12}+b_{13}(x+1)^{13}+b_{14}(x+1)^{14}$ and $b_7,b_9,b_{11},b_{13},b_{14}\in \mathbb{F}_{2^m}$;

\item{$\diamond$}
 $\langle (x+1)^8b+u(x+1)^7, (x+1)^9\rangle$ where $b\in \mathbb{F}_{2^m}$;

\item{$\diamond$}
 $\langle (x+1)^7b(x)+u(x+1)^6, (x+1)^{10}\rangle$, where $b(x)=b_1(x+1)+b_2(x+1)^2$ and $b_1,b_2\in \mathbb{F}_{2^m}$;

\item{$\diamond$}
 $\langle (x+1)^6b(x)+u(x+1)^5, (x+1)^{11}\rangle$, where $b(x)=b_3(x+1)^3+b_4(x+1)^4$ and $b_3,b_4\in \mathbb{F}_{2^m}$;

\item{$\diamond$}
 $\langle (x+1)^5b(x)+u(x+1)^4, (x+1)^{12}\rangle$, where $b(x)=b_3(x+1)^3+b_5(x+1)^5+b_6(x+1)^6$ and $b_3,b_5,b_6\in \mathbb{F}_{2^m}$;

\item{$\diamond$}
 $\langle (x+1)^4b(x)+u(x+1)^3, (x+1)^{13}\rangle$, where $b(x)=b_5(x+1)^5+b_5(x+1)^6+b_7(x+1)^7+b_8(x+1)^8$ and $b_5,b_7,b_8\in \mathbb{F}_{2^m}$;

\item{$\diamond$}
 $\langle (x+1)^3b(x)+u(x+1)^2, (x+1)^{14}\rangle$, where $b(x)=b_5(x+1)^5+b_5(x+1)^6+b_7(x+1)^7+b_9(x+1)^9+b_{10}(x+1)^{10}$ and $b_5,b_7,b_9,b_{10}\in \mathbb{F}_{2^m}$;

\item{$\diamond$}
  $\langle (x+1)^2b(x)+u(x+1), (x+1)^{15}\rangle$, where $b(x)=b_7(x+1)^7+b_9(x+1)^9+b_9(x+1)^{10}+b_{11}(x+1)^{11}+b_{12}(x+1)^{12}$ and $b_7,b_9,b_{11},b_{12}\in \mathbb{F}_{2^m}$.
\end{description}

\vskip 3mm\noindent
  {\bf Remark.} (i) In Corollary 22(ii) of [5], the authors claimed that the number
of Euclidean self-dual cyclic codes of length $2^s$ over $\mathbb{F}_{2^k}+u\mathbb{F}_{2^k}$ ($u^2=0$) is
$${\rm NE}(2^k,2^s)=\left\{\begin{array}{ll} 1+2^k & {\rm if} \ s=1,
 \cr 1+2^k+(2^k)^2 & {\rm if} \ s=2,
 \cr 1+2^k+2(2^k)^2\cdot\frac{(2^k)^{2^{s-2}-1}-1}{2^k-1} & {\rm if} \ s\geq 3.\end{array}\right.$$
It is obvious that the above formula missed the term $(2^k)^{2^{s-2}+1}$ when $s\geq 3$.
  For examples, there are $1+2+2\cdot 2^2+2^3=19$ Euclidean self-dual cyclic codes of length $2^3$ over $\mathbb{F}_{2}+u\mathbb{F}_{2}$, not just $1+2+2\cdot 2^2=11$ Euclidean self-dual cyclic codes of length $2^3$ over $\mathbb{F}_{2}+u\mathbb{F}_{2}$.

\par
  (ii) From self-dual cyclic codes over $\mathbb{F}_{2^m}+u\mathbb{F}_{2^m}$ of length $2^s$ and by the isomorphism $\phi$ of $\mathbb{F}_{2^m}$-linear spaces from $\frac{(\mathbb{F}_{2^m}+u\mathbb{F}_{2^m})[x]}{\langle x^{2^s}-1\rangle}$ onto $\mathbb{F}_{2^m}^{2^{s+1}}$ defined in Section 1, we obtain $N_s=1+2^m+2(2^m)^2\cdot\frac{(2^m)^{2^{s-2}-1}-1}{2^m-1}+(2^m)^{2^{s-2}+1}$ self-dual and
$2$-quasi-cyclic codes over $\mathbb{F}_{2^m}$ of length $2^{s+1}$ for any $s\geq 3$.

\section{Conclusions and further research}
\noindent
We have given an explicit representation for self-dual cyclic codes of length $2^s$
over the finite chain ring $\mathbb{F}_{2^m}
+u\mathbb{F}_{2^m}$ ($u^2=0$) and a clear mass formula to count the number
of these codes, by a new way different from that used in [5]. Especially, we provide an effective method to list precisely all distinct self-dual cyclic codes of length $2^s$
over $\mathbb{F}_{2^m}+u\mathbb{F}_{2^m}$
by use of properties for Kronecker product of matrices and calculation of linear equations over $\mathbb{F}_{2^m}$.

\par
   Giving an explicit
representation and enumeration for self-dual cyclic codes
over $\mathbb{F}_{2^m}+u\mathbb{F}_{2^m}$ for arbitrary even length and obtaining some bounds for minimal distance such as BCH-like of a self-dual cyclic code over the ring $\mathbb{F}_{2^m}+u\mathbb{F}_{2^m}$ by just looking at the representation of such codes are future topics of interest.

\section*{Acknowledgments}

\noindent
Part of this work was
done when Yonglin Cao was visiting Chern Institute of Mathematics, Nankai
University, Tianjin, China. Yonglin Cao would like to thank the institution
for the kind hospitality. This research is supported in part by the National
Natural Science Foundation of China (Grant Nos. 11671235, 11801324), the Shandong Provincial Natural Science Foundation,
China (Grant No. ZR2018BA007) and the Scientific Research Fund of Hunan
Provincial Key Laboratory of Mathematical Modeling and Analysis in
Engineering (No. 2018MMAEZD04). H.Q. Dinh is
grateful for the Centre of Excellence in Econometrics, Chiang Mai University,
Thailand, for partial financial support.

\section*{Appendix: Calculation for the set $\mathcal{S}_l$, $1\leq l\leq 16$}

   ($\dag$) First, we determine the matrices $M_l$ by Theorem 1(ii), for all $l=1,2,\ldots,8$.

   $\triangleright$ $M_2=I_{2}+G_{2}=\left(\begin{array}{cc} 0 & 0 \cr 1 & 0\end{array}\right)$ and $M_1=(0)=0$.

\par
   $\triangleright$ $M_4=I_{4}+G_{4}=\left(\begin{array}{cc|cc} 0 & 0 & 0 & 0 \cr 1 & 0 & 0 & 0 \cr\hline
 1 & 0 & 0 & 0 \cr 1 & 1 & 1 & 0 \end{array}\right)$ and
$M_3=\left(\begin{array}{cc|c} 0 & 0 & 0 \cr 1 & 0 & 0 \cr\hline
 1 & 0 & 0 \end{array}\right)$.

\par
   $\triangleright$ $M_8=I_{8}+G_{8}=\left(\begin{array}{cccc|cccc} 0 & 0 & 0 & 0 & 0 & 0 & 0 & 0\cr 1 & 0 & 0 & 0 & 0 & 0 & 0 & 0\cr
 1 & 0 & 0 & 0 & 0 & 0 & 0 & 0\cr 1 & 1 & 1 & 0 & 0 & 0 & 0 & 0\cr\hline
 1 & 0 & 0 & 0 & 0 & 0 & 0 & 0 \cr 1 &  1 & 0 & 0  & 1 & 0 & 0 & 0\cr
 1 & 0 & 1 & 0 & 1 & 0 & 0 & 0 \cr 1 & 1 & 1 & 1 & 1 &  1 & 1 & 0
 \end{array}\right)$,

 $M_5=\left(\begin{array}{cccc|c} 0 & 0 & 0 & 0 & 0 \cr 1 & 0 & 0 & 0 & 0 \cr
 1 & 0 & 0 & 0 & 0 \cr 1 & 1 & 1 & 0 & 0 \cr\hline
 1 & 0 & 0 & 0 & 0  \end{array}\right)$,
$M_6=\left(\begin{array}{cccc|cc} 0 & 0 & 0 & 0 & 0 & 0 \cr 1 & 0 & 0 & 0 & 0 & 0 \cr
 1 & 0 & 0 & 0 & 0 & 0 \cr 1 & 1 & 1 & 0 & 0 & 0 \cr\hline
 1 & 0 & 0 & 0 & 0 & 0 \cr 1 & 1 & 0 & 0  & 1 & 0
 \end{array}\right)$

  and
$M_{7}=\left(\begin{array}{cccc|ccc} 0 & 0 & 0 & 0 & 0 & 0 & 0 \cr 1 & 0 & 0 & 0 & 0 & 0 & 0 \cr
 1 & 0 & 0 & 0 & 0 & 0 & 0 \cr 1 & 1 & 1 & 0 & 0 & 0 & 0 \cr\hline
 1 & 0 & 0 & 0 & 0 & 0 & 0 \cr 1 & 1 & 0 & 0 & 1 & 0 & 0\cr
 1 & 0 & 1 & 0 & 1 & 0 & 0
 \end{array}\right)$.

\vskip 2mm\par
  ($\ddag$) Then we determine $\mathcal{S}_l$ by by Theorem 1(iii), for all $l=1,2,3,\ldots,16$.
    Obviously, we have

\begin{description}
\item{$\diamond$}
     $\mathcal{S}_1=\mathbb{F}_{2^m}$, and $\mathcal{S}_2=\{(0,b_1)^{{\rm tr}}\mid
b_1\in \mathbb{F}_{2^m}\}$ with $|\mathcal{S}_2|=2^m$.
\end{description}

\par
   ($\ddag$-1) Let $B_2=(0,b_1)^{{\rm tr}} \in \mathcal{S}_2$ arbitrary.
   Solving the linear equations $M_1B_3^{(3,3)}=B_2^{(1,1)}=0$, i.e. $0\cdot b_2=0$, we obtain $B_3^{(3,3)}=b_2\in \mathbb{F}_{2^m}$. Hence
\begin{description}
\item{$\diamond$}
 $\mathcal{S}_3=\{B_3=\left(\begin{array}{c}B_2\cr B_3^{(3,3)}\end{array}\right)=\left(0,b_1 | \ b_2\right)^{{\rm tr}} \mid b_1,b_2\in \mathbb{F}_{2^m}\}$.
\end{description}

\par
  Solving $M_2B_4^{(3,4)}=B_2^{(1,2)}=B_2$, i.e.
  $\left(\begin{array}{ccc} 0 & 0 \cr 1 & 0\end{array}\right)\left(\begin{array}{c} b_2 \cr b_3\end{array}\right)
  =\left(\begin{array}{c}0\cr b_1\end{array}\right)  $, we obtain
$B_4^{(3,4)}=(b_1,b_3)^{{\rm tr}}$ where $b_1, b_3\in \mathbb{F}_{2^m}$. Hence

\begin{description}
\item{$\diamond$}
 $\mathcal{S}_4=\{B_4=\left(\begin{array}{c}B_2\cr B_4^{(3,4)}\end{array}\right)=
\left(0,b_1 | \ b_1,b_3\right)^{{\rm tr}}\mid b_1,b_3\in \mathbb{F}_{2^m}\}$.
\end{description}

\par
   ($\ddag$-2) Let $B_4=(0,b_1,b_1,b_3)^{{\rm tr}}\in \mathcal{S}_4$ arbitrary.
   Solving $M_1B_5^{(5,5)}=B_4^{(1,1)}$, i.e. $0\cdot b_4=0$, we obtain $B_5^{(5,5)}=b_4\in \mathbb{F}_{2^m}$. Hence

\begin{description}
\item{$\diamond$}
  $\mathcal{S}_5=\{B_5=\left(\begin{array}{c}B_4\cr B_5^{(5,5)}\end{array}\right)
 =\left(0,b_1,b_1,b_3  | b_4\right)^{{\rm tr}}\mid b_1,b_3,b_4\in \mathbb{F}_{2^m}\}$.
\end{description}

\par
  Solving $M_2B_6^{(5,6)}=B_4^{(1,2)}$, i.e.
  $\left(\begin{array}{cc} 0 & 0 \cr 1 & 0\end{array}\right)\left(\begin{array}{c} b_4 \cr b_5\end{array}\right)
  =\left(\begin{array}{c}0\cr b_1\end{array}\right)$, we obtain
  $B_6^{(5,6)}=(b_1, b_5)^{{\rm tr}}$ where $b_1, b_5\in \mathbb{F}_{2^m}$. Hence

\begin{description}
\item{$\diamond$}
  $\mathcal{S}_6=\{B_6=\left(\begin{array}{c}B_4\cr B_6^{(5,6)}\end{array}\right)
  =\left(0,b_1,b_1,b_3 | \ b_1, b_5\right)^{{\rm tr}}\mid b_1,b_3,b_5\in \mathbb{F}_{2^m}\}$.
\end{description}

\par
  Solving $M_3B_7^{(5,7)}=B_4^{(1,3)}$, i.e.
  $\left(\begin{array}{ccc} 0 & 0 & 0\cr 1 & 0 & 0 \cr 1 & 0 & 0 \end{array}\right)\left(\begin{array}{c} b_4 \cr b_5\cr b_6\end{array}\right)
  =\left(\begin{array}{c}0\cr b_1\cr b_1\end{array}\right)$, we obtain
  $B_7^{(5,7)}=(b_1, b_5, b_6)^{{\rm tr}}$ where $b_1, b_5,b_6\in \mathbb{F}_{2^m}$.
  Hence

\begin{description}
\item{$\diamond$}
 $\mathcal{S}_7=\{B_7=\left(\begin{array}{c}B_4\cr B_7^{(5,7)}\end{array}\right)
  =\left(0,b_1,b_1,b_3 | \ b_1, b_5,b_6\right)^{{\rm tr}}\mid b_1,b_3,b_5,b_6\in \mathbb{F}_{2^m}\}.$
\end{description}

\par
  Solving $M_4B_8^{(5,8)}=B_4^{(1,4)}=B_4$, i.e.
  $\left(\begin{array}{cc|cc} 0 & 0 & 0 & 0\cr 1 & 0 & 0 & 0\cr\hline 1 & 0 & 0 & 0\cr 1 & 1 & 1& 0\end{array}\right)\left(\begin{array}{c} b_4 \cr b_5\cr b_6 \cr b_7\end{array}\right)
  =\left(\begin{array}{c}0\cr b_1\cr b_1\cr b_3\end{array}\right)$, we obtain
  $B_8^{(5,8)}=(b_1, b_5, b_1+b_3+b_5, b_7)^{{\rm tr}}$ where $b_1, b_5,b_7\in \mathbb{F}_{2^m}$. Hence

\begin{description}
\item{$\diamond$}
 $\mathcal{S}_8=\{B_8=\left(\begin{array}{c}B_4\cr B_8^{(5,8)}\end{array}\right)
 =\left(0,b_1,b_1,b_3 | \ b_1, b_5,b_1+b_3+b_5, b_7\right)^{{\rm tr}}\mid b_1,b_3,b_5$, $b_7\in \mathbb{F}_{2^m}\}.$
\end{description}

\par
   ($\ddag$-3) Let $B_8=(0,b_1,b_1,b_3, b_1, b_5,b_1+b_3+b_5, b_7)^{{\rm tr}} \in \mathcal{S}_8$ arbitrary.
Solving $M_1B_{9}^{(9,9)}=B_8^{(1,1)}$, i.e. $0\cdot b_8=0$, we obtain $B_{9}^{(9,9)}=b_8\in \mathbb{F}_{2^m}$. Hence

\begin{description}
\item{$\diamond$}
  $\mathcal{S}_9=\{B_9=\left(\begin{array}{c}B_8\cr B_{9}^{(9,9)}\end{array}\right)
  =\left(0,b_1,b_1,b_3, b_1, b_5,b_1+b_3+b_5, b_7  | \ b_8\right)^{{\rm tr}}\mid b_1,b_3$, $b_5,b_7,b_8\in \mathbb{F}_{2^m}\}$.
\end{description}

\par
  Solving $M_2B_{10}^{(9,10)}=B_8^{(1,2)}$, i.e.
  $\left(\begin{array}{cc} 0 & 0 \cr 1 & 0\end{array}\right)\left(\begin{array}{c} b_8 \cr b_9\end{array}\right)
  =\left(\begin{array}{c}0\cr b_1\end{array}\right)$, we obtain
  $B_{10}^{(9,10)}=(b_1, b_9)^{{\rm tr}}$ where $b_1, b_9\in \mathbb{F}_{2^m}$.
  Hence

\begin{description}
\item{$\diamond$}
 $\mathcal{S}_{10}=\{B_{10}=\left(\begin{array}{c}B_8\cr B_{10}^{(9,10)}\end{array}\right)
  =\left(0,b_1,b_1,b_3, b_1, b_5,b_1+b_3+b_5, b_7  | \ b_1, b_9\right)^{{\rm tr}}\mid b_1,b_3,b_5, b_7,b_9\in \mathbb{F}_{2^m}\}$.
\end{description}

\par
  Solving $M_3B_{11}^{(9,11)}=B_8^{(1,3)}$, i.e.
  $\left(\begin{array}{ccc} 0 & 0 & 0\cr 1 & 0 & 0 \cr 1 & 0 & 0 \end{array}\right)\left(\begin{array}{c} b_8 \cr b_9\cr b_{10}\end{array}\right)
  =\left(\begin{array}{c}0\cr b_1\cr b_1\end{array}\right)$, we obtain
  $B_{11}^{(9,11)}=(b_1, b_9,b_{10})^{{\rm tr}}$ where $b_1, b_9,b_{10}\in \mathbb{F}_{2^m}$. Hence

\begin{description}
\item{$\diamond$}
  $\mathcal{S}_{11}=\{B_{11}=\left(\begin{array}{c}B_8\cr B_{11}^{(9,11)}\end{array}\right)=
   \left(0,b_1,b_1,b_3, b_1, b_5,b_1+b_3+b_5, b_7  | \ b_1, b_9, b_{10}\right)^{{\rm tr}}$
   $\mid b_1,b_3,b_7,b_5,b_9,b_{10}\in \mathbb{F}_{2^m}\}$.
\end{description}

\par
  Solving $M_4B_{12}^{(9,12)}=B_8^{(1,4)}=B_4$, i.e.
  $\left(\begin{array}{cc|cc} 0 & 0 & 0 & 0\cr 1 & 0 & 0 & 0\cr\hline 1 & 0 & 0 & 0\cr 1 & 1 & 1& 0\end{array}\right)\left(\begin{array}{c} b_8 \cr b_{9}\cr b_{10} \cr b_{11}\end{array}\right)
  =B_4$, we obtain
  $B_{12}^{(9,12)}=(b_1, b_9, b_1+b_3+b_9, b_{11})^{{\rm tr}}$ where $b_1,b_3, b_9,b_{11}\in \mathbb{F}_{2^m}$. Hence

\begin{description}
\item{$\diamond$}
 $\mathcal{S}_{12}=\{B_{12}=\left(\begin{array}{c}B_8\cr B_{12}^{(9,12)}\end{array}\right)
=\left(0,b_1,b_1,b_3, b_1, b_5,b_1+b_3+b_5, b_7  | \ b_1, b_9, b_1\right.$ $\left.+b_3+b_9, b_{11}\right)^{{\rm tr}}
\mid b_1,b_3,b_5,b_7, b_9,b_{11}\in \mathbb{F}_{2^m}\}$.
\end{description}

\par
  Solving $M_5B_{13}^{(9,13)}=B_8^{(1,5)}$, i.e.
  $\left(\begin{array}{c|c} M_4 & 0\cr\hline (1,0,0,0) & 0\end{array}\right)\left(\begin{array}{c} B_{12}^{(9,12)}
  \cr\hline b_{12}\end{array}\right)
  =\left(\begin{array}{c} B_4\cr\hline b_1\end{array}\right)$, we obtain
  $B_{13}^{(9,13)}=\left(\begin{array}{c} B_{12}^{(9,12)}
  \cr\hline b_{12}\end{array}\right)$ where $b_1,b_3, b_9,b_{11}\in \mathbb{F}_{2^m}$. Hence

\begin{description}
\item{$\diamond$}
 $\mathcal{S}_{13}=\{B_{13}=\left(\begin{array}{c}B_8\cr B_{13}^{(9,13)}\end{array}\right)
=\left(\begin{array}{c}B_{12}\cr\hline b_{12}\end{array}\right)
\mid B_{12}\in \mathcal{S}_{12}, \ b_{12}\in \mathbb{F}_{2^m}\}$.
\end{description}

\par
  We consider the matrix equation $M_6B_{14}^{(9,14)}=B_8^{(1,6)}$, i.e.,
  $$\left(\begin{array}{c|c} M_4 & 0\cr\hline (G_2,0) & M_2\end{array}\right)\left(\begin{array}{c} B_{12}^{(9,12)}
  \cr\hline b_{12}\cr b_{13}\end{array}\right)
  =\left(\begin{array}{c} B_4\cr\hline b_1\cr b_5\end{array}\right).$$
This is equivalent to $M_4B_{12}^{(9,12)}=B_4$ and
$$\left(\begin{array}{cc} 1 & 0 \cr 1 & 1\end{array}\right)\left(\begin{array}{c}b_1 \cr b_9\end{array}\right)
+M_2\left(\begin{array}{c}b_{12} \cr b_{13}\end{array}\right)=\left(\begin{array}{c}b_1 \cr b_5\end{array}\right).$$
From this we deduce that $b_{12}=b_1+b_5+b_9$ and $b_{13}\in\mathbb{F}_{2^m}$ arbitrary. Hence

\begin{description}
\item{$\diamond$}
 $\mathcal{S}_{14}=\{B_{14}=\left(\begin{array}{c}B_8\cr B_{14}^{(9,14)}\end{array}\right)
=\left(\begin{array}{c}B_{12}\cr\hline b_1+b_5+b_9\cr b_{13}\end{array}\right)
\mid b_1,b_3,b_5,b_7, b_9,b_{11},b_{13}$ $\in \mathbb{F}_{2^m}\}$ in which
$B_{12}\in \mathcal{S}_{12}$.
\end{description}

\par
  We consider the matrix equation $M_7B_{15}^{(9,15)}=B_8^{(1,7)}$, i.e.,
  $$\left(\begin{array}{c|c} M_4 & 0\cr\hline (G_2,0) & (M_2,0) \cr (1,0,1,0) & (1,0,0)\end{array}\right)\left(\begin{array}{c} B_{12}^{(9,12)}
  \cr\hline b_{12}\cr b_{13}\cr b_{14}\end{array}\right)
  =\left(\begin{array}{c} B_4\cr\hline b_1\cr b_5\cr b_1+b_3+b_5\end{array}\right).$$
This is equivalent to $M_4B_{12}^{(9,12)}=B_4$ and
$$\left(\begin{array}{ccc} 1 & 0 & 0\cr 1 & 1 & 0 \cr 1 & 0 & 1\end{array}\right)
\left(\begin{array}{c}b_1 \cr b_9 \cr b_1+b_3+b_9\end{array}\right)
+\left(\begin{array}{c} M_2\cr (1,0)\end{array}\right)\left(\begin{array}{c}b_{12} \cr b_{13}\end{array}\right)=\left(\begin{array}{c}b_1 \cr b_5\cr b_1+b_3+b_5\end{array}\right).$$
From this we deduce that $b_{12}=b_1+b_5+b_9$ and $b_{13},b_{14}\in\mathbb{F}_{2^m}$ arbitrary. Hence

\begin{description}
\item{$\diamond$}
 $\mathcal{S}_{15}=\{B_{15}=\left(\begin{array}{c}B_8\cr B_{15}^{(9,15)}\end{array}\right)
=\left(\begin{array}{c}B_{12}\cr\hline b_1+b_5+b_9\cr b_{13}\cr b_{14}\end{array}\right)
\mid b_1,b_3,b_5,b_7, b_9,b_{11},b_{13}$, $b_{14}\in \mathbb{F}_{2^m}\}$ in which
$B_{12}\in \mathcal{S}_{12}$.
\end{description}

\par
  We consider the matrix equation $M_8B_{16}^{(9,16)}=B_8^{(1,8)}=B_8$, i.e.,
  $$\left(\begin{array}{c|c} M_4 & 0\cr\hline G_4 & M_4\end{array}\right)\left(\begin{array}{c} B_{12}^{(9,12)}
  \cr\hline b_{12}\cr b_{13}\cr b_{14}\cr b_{15}\end{array}\right)
  =\left(\begin{array}{c} B_4\cr\hline b_1\cr b_5\cr b_1+b_3+b_5\cr b_7\end{array}\right).$$
This is equivalent to $M_4B_{12}^{(9,12)}=B_4$ and
$$G_4
\left(\begin{array}{c}b_1 \cr b_9 \cr b_1+b_3+b_9 \cr b_{11}\end{array}\right)
+M_4\left(\begin{array}{c}b_{12} \cr b_{13}\cr b_{14}\cr b_{15}\end{array}\right)=\left(\begin{array}{c}b_1 \cr b_5\cr b_1+b_3+b_5\cr b_7\end{array}\right).$$
From this we deduce that $b_{12}=b_1+b_5+b_9$, $b_{14}=\sum_{i=1}^7b_{2i-1}$ and $b_{13},b_{15}\in\mathbb{F}_{2^m}$ arbitrary. Hence

\begin{description}
\item{$\diamond$}
  $\mathcal{S}_{16}=\{B_{16}=\left(\begin{array}{c}B_8\cr B_{16}^{(9,16)}\end{array}\right)
=\left(\begin{array}{c}B_{12}\cr\hline b_1+b_5+b_9\cr b_{13}\cr \sum_{i=1}^7b_{2i-1}\cr b_{15}\end{array}\right)
\mid b_1,b_3,b_5,b_7, b_9,b_{11},b_{13}$, $b_{15}\in \mathbb{F}_{2^m}\}$ in which
$B_{12}\in \mathcal{S}_{12}$.
\end{description}



\begin{thebibliography}{99}

\bibitem{s1} T. Abualrub, I. Siap, Constacyclic codes over $\mathbb{F}_2+u\mathbb{F}_2$,
J. Franklin Inst. {\bf 346} (2009), 520--529.

\bibitem{s2} M. C. V. Amerra, F. R. Nemenzo, On $(1-u)$-cyclic codes over $\mathbb{F}_{p^k}+u\mathbb{F}_{p^k}$,
Appl. Math. Lett. {\bf 21} (2008), 1129--1133.

\bibitem{s3} K. Betsumiya, S. Ling, and F. R. Nemenzo, Type II codes
over $\mathbb{F}_{2^m}+u\mathbb{F}_{2^m}$, Discrete Math. {\bf 275} (2004), no. 1-3, 43--65.

\bibitem{s4} A. Bonnecaze, P. Udaya, Cyclic codes and self-dual codes over $\mathbb{F}_2+u\mathbb{F}_2$,
IEEE Trans. Inform. Theory {\bf 45} (1999), 1250--1255.

\bibitem{s5} P. Choosuwan, S. Jitman, P. Udomkavanich, Self-dual abelian codes in some nonprincipal ideal
group algebras, Mathematical Problems in Engineering, Vol.2016, Article ID 9020173, 12 pages.

\bibitem{s6}  Y. Cao, Y. Cao, H. Q. Dinh, F-W. Fu, J. Gao, S. Sriboonchitta, Constacyclic codes of length $np^s$ over $\mathbb{F}_{p^m} + u\mathbb{F}_{p^m} $, Adv. Math. Commun. {\bf 12} (2018), 231--262.

\bibitem{s7} B. Chen, H. Q. Dinh, H. Liu , L. Wang, Constacyclic codes of length $2p^s$ over $\mathbb{F}_{p^m}+u \mathbb{F}_{p^m}$,
Finite Fields Appl. {\bf 37} (2016), 108--130.

\bibitem{s8} H. Q. Dinh, Constacyclic codes of length $2^s$ over
Galois extension rings of $\mathbb{F}_2+u\mathbb{F}_2$, IEEE Trans.
Inform. Theory {\bf 55} (2009), 1730--1740.

\bibitem{s9} H. Q. Dinh, Constacyclic codes of length $p^s$ over
$\mathbb{F}_{p^m}+u \mathbb{F}_{p^m}$, J. Algebra, {\bf 324} (2010),
940--950.

\bibitem{s10} H. Q. Dinh, L. Wang, S. Zhu, Negacyclic codes of length $2p^s$ over $\mathbb{F}_{p^m}+u \mathbb{F}_{p^m}$, Finite Fields Appl., {\bf 31} (2015), 178-201.

\bibitem{s11} H. Q. Dinh, S. Dhompongsa, and S. Sriboonchitta, On constacyclic codes of length
$4p^s$ over $\mathbb{F}_{p^m}+u \mathbb{F}_{p^m}$, Discrete Math. {\bf 340} (2017), 832--849.

\bibitem{s12} H. Q. Dinh, A. Sharma, S. Rani, and S. Sriboonchitta, Cyclic and negacyclic codes of length
$4p^s$ over $\mathbb{F}_{p^m}+u \mathbb{F}_{p^m}$, J. Algebra Appl. Vol. 17, No. 9 (2018) 1850173 (22 pages).

\bibitem{s13} H. Q. Dinh, S. R. L\'{o}pez-Permouth, Cyclic and negacyclic codes over finite chain rings,
IEEE Trans. Inform. Theory {\bf 50} (2004), 1728--1744.

\bibitem{s14} S. T. Dougherty, P. Gaborit, M. Harada, P. Sole, Type II codes over $\mathbb{F}_2+u\mathbb{F}_2$,
IEEE Trans. Inform. Theory {\bf 45} (1999), 32--45.

\bibitem{s15}  S. T. Dougherty,  J-L. Kim,  H. Kulosman,  H. Liu: Self-dual
codes over commutative Frobenius rings, Finite Fields Appl. {\bf 16} (2010), 14--26.

\bibitem{s16} T. A. Gulliver, M. Harada, Construction of optimal Type IV self-dual codes over $\mathbb{F}_2+u\mathbb{F}_2$,
IEEE Trans. Inform. Theory {\bf 45} (1999), 2520--2521.

\bibitem{s17} W. C. Huffman, On the decompostion of self-dual codes over $\mathbb{F}_2+u\mathbb{F}_2$
with an automorphism of odd prime number, Finite Fields Appl. {\bf 13} (2007), 682--712.

\bibitem{s18} H.M. Kiah, K. H. Leung, and S. Ling, A note on cyclic codes
over ${\rm GR}(p^2, m)$ of length $p^k$, Des., Codes and Cryptog., {\bf 63}, no. 1, (2012), 105--112.

\bibitem{s19} J. F. Qian, L. N. Zhang, S. Zhu, $(1+u)$-constacyclic and cyclic codes over $\mathbb{F}_2+u\mathbb{F}_2$,
Appl. Math. Lett. {\bf 19} (2006), 820--823.

\end{thebibliography}
\end{document}